# Rate-dependent effects of lidocaine on cardiac dynamics: Development and analysis of a low-dimensional drug-channel interaction model

Steffen S. Docken[1,2,#a], Colleen E. Clancy[2], Timothy J. Lewis[1*]

[1]Department of Mathematics, University of California Davis, Davis, CA, United States of America

[2]Department of Physiology and Membrane Biology, School of Medicine, University of California Davis, Davis, CA, United States of America

[#a]Current Address: Kirby Institute for Infection and Immunity, University of New South Wales, Sydney, NSW, Australia

* Corresponding author

E-mail: tjlewis@ucdavis.edu (TJL)

## Abstract

State-dependent sodium channel blockers are often prescribed to treat cardiac arrhythmias, but many sodium channel blockers are known to have pro-arrhythmic side effects. While the anti and proarrhythmic potential of a sodium channel blocker is thought to depend on the characteristics of its rate-dependent block, the mechanisms linking these two attributes are unclear. Furthermore, how specific properties of rate-dependent block arise from the binding kinetics of a particular drug is poorly understood. Here, we examine the rate-dependent effects of the sodium channel blocker lidocaine by constructing and analyzing a novel drug-channel interaction model. First, we identify the predominant mode of lidocaine binding in a 24 variable Markov model for lidocaine-sodium channel interaction by Moreno et al. Specifically, we find that (1) the vast majority of lidocaine bound to sodium channels is in the neutral form, i.e., the binding of charged lidocaine to sodium channels is negligible, and (2) neutral lidocaine binds almost exclusively to inactivated channels and, upon binding, immobilizes channels in the inactivated state. We then develop a novel 3-variable lidocaine-sodium channel interaction model that incorporates only the predominant mode of drug binding. Our low-dimensional model replicates the extensive voltage-clamp data used to parameterize the Moreno et al. model. Furthermore, the effects of lidocaine on action potential upstroke velocity and conduction velocity in our model are similar to those predicted by the Moreno et al. model. By exploiting the low-dimensionality of our model, we derive an algebraic expression for level of rate-dependent block as a function of pacing frequency, restitution properties, diastolic and plateau potentials, and drug binding rate constants. Our model predicts that the level of rate-dependent block is sensitive to alterations in restitution properties and increases in diastolic potential, but it is insensitive to variations in the shape of the action potential waveform and lidocaine binding rates.

## Author Summary

Cardiac arrhythmias are often treated with drugs that block and alter the kinetics of membrane sodium channels. However, different drugs interact with sodium channels in different ways, and the complexity of the drug-channel interactions make it difficult to predict whether a particular sodium channel blocker will reduce or increase the probability of cardiac arrhythmias. Here, we characterize the binding kinetics and effects on electrical signal propagation of the antiarrhythmic drug lidocaine, which is an archetypical example of a safe sodium channel blocker. Through analysis of a high-dimensional biophysically-detailed model of lidocaine-sodium channel interaction, we identify the predominant lidocaine binding pathway. We then incorporate only the key features of the predominant binding pathway into a novel low-dimensional model of lidocaine-sodium channel interaction. Our analysis of the low-dimensional model characterizes how the key binding properties of lidocaine affect electrical signal generation and propagation in the heart, and therefore our results are a step towards understanding the features that differentiate pro- and antiarrhythmic sodium channel blockers.

## 1. Introduction

Class I antiarrhythmic drugs bind to and block the fast sodium ($Na^+$) channels that are responsible for the upstroke and propagation of the cardiac action potential. $Na^+$ channel blockers are thought to exert their antiarrhythmic effects by reducing the excitability of cardiac tissue in a rate-dependent manner, thereby preventing aberrant spontaneous action potentials. However, blocking the $Na^+$ current also slows action potential conduction, which can facilitate the onset of reentrant arrhythmias, especially at high heart rates [1, 2]. Indeed, many $Na^+$ channel blockers that are prescribed as antiarrhythmic drugs, such as flecainide, can increase the propensity of ventricular arrhythmias [3, 4], whereas other $Na^+$ channel blockers, such as lidocaine, are considered to be safe [5]. The mechanisms that make certain $Na^+$ channel blockers antiarrhythmic and others pro-arrhythmic remain unclear.

To determine the arrhythmic potential of $Na^+$ channel blockers, biophysically detailed models of $Na^+$ channel-drug interactions have been constructed using a detailed Markov model framework [4, 6-8]. Such models accurately reproduce voltage-clamp data that characterize the modulatory effects of drugs on channel kinetics [9-13], but come at the cost of being complex, consisting of up to 26 dynamic variables. While these high-dimensional models have been able to help determine if specific drugs prevent or exacerbate arrhythmia, their complexity impedes the ability to identify the fundamental features of drug-channel interactions that lead to pro-arrhythmic effects. This in turn makes it difficult to generalize knowledge of well-studied $Na^+$ channel blockers, such as lidocaine or flecainide, to new drugs. Low dimensional or "minimal" models that contain only the key features of drug-ion channel interactions can be advantageous in identifying the fundamental mechanisms that underlie the rate-dependent block and resulting pro- or antiarrhythmic effects of some $Na^+$ channel blockers, as such mechanisms would be brought to the foreground due to the simplicity of such models.

In this paper, we construct and analyze a novel low-dimensional lidocaine-$Na^+$ channel interaction model that consists of 3 dynamic variables. We first analyze a detailed Markov model for lidocaine-$Na^+$ channel interaction introduced by Moreno et al. [4] and identify the predominant mode of binding that defines lidocaine's effects on the $Na^+$ current. We then base the structure of our low-dimensional model on these essential features. The low-dimensional model is fit to voltage-clamp data that was used to parameterize the Moreno et al. model. We observe that our low-dimensional model and the high-dimensional Moreno et al. model reproduce the experimental data to a similar degree for physiological conditions. Additionally, we show that our model and the Moreno et al. model predict similar rate-dependent effects on action potential upstroke and conduction velocity. Finally, we utilize the simple structure of our low-dimensional model to understand the rate-dependent effects of lidocaine and how these effects are influenced by various physiological properties (e.g., the action potential duration restitution curve and the diastolic transmembrane potential).

## 2. Models of fast $Na^+$ current conductance

Cardiac $Na^+$ channels are generally modeled as variable Ohmic resistors. The $Na^+$ current is the product of the $Na^+$ conductance ($\bar{g}_{Na} f_{open}$) and the driving force ($V - E_{Na}$),

$$I_{Na} = \bar{g}_{Na} f_{open} (V - E_{Na}), \tag{1}$$

where $V$ is the transmembrane potential, $E_{Na}$ is the Na$^+$ reversal potential, $\bar{g}_{Na}$ is the maximal conductance (i.e., the conductance when all channels are open), and $f_{open}$ is the fraction of channels that are in an open and unblocked state. In this section, we describe and compare two mathematical models for the dynamics of $f_{open}$ in the presence of lidocaine: A detailed Markov model by Moreno et al. [4], and a novel low-dimensional model with generalized Hodgkin-Huxley formalism that we develop here.

**2.1. The Moreno et al. model: A detailed Markov model**

The conformational state diagram of the drug-free Moreno et al. model for Na$^+$ channel dynamics is displayed in Fig 1A. Channels can be in an open state (*O*), three closed states (*C3*, *C2*, and *C1*), three fast-inactivated states (*IC3*, *IC2*, and *IF*), or a slow-inactivated state (*IS*) [4]. Conformational state transitions are indicated by arrows, with the corresponding voltage-dependent rate constants of each transition indicated by $\alpha$s and $\beta$s.

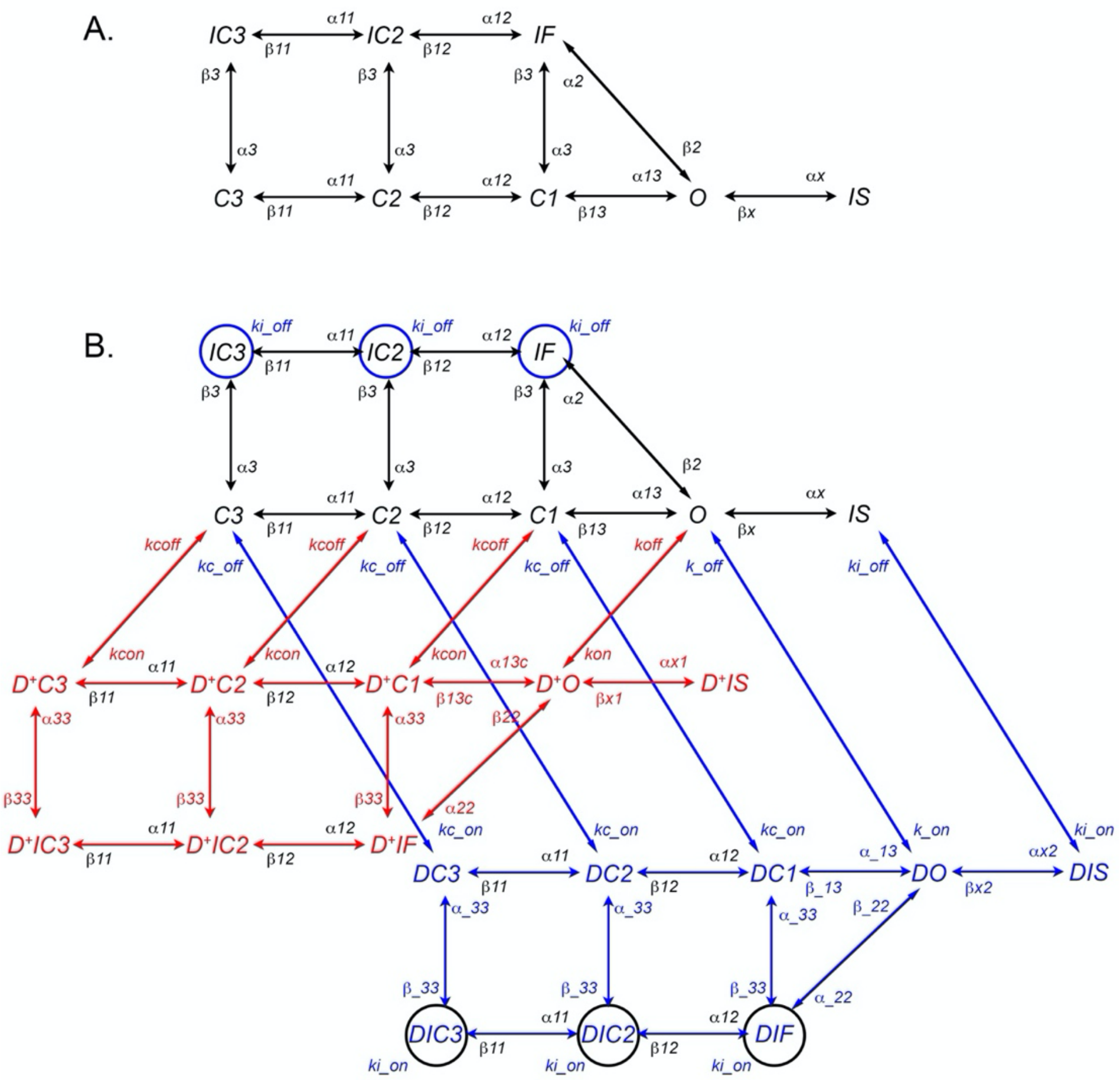

**Fig 1. Moreno et al. [4] model of Na+ channel-lidocaine interactions.** *(A) The drug-free Na+ channel model. The $O$ state represents the conducting state, while $C3$, $C2$, and $C1$ correspond to 3 closed states. The $IC3$, $IC2$, and $IF$ states represent conformational states in which the "fast" inactivation gate is closed, and the $IS$ state represents a state in which a "slow" inactivation gate is closed. Arrows indicate possible conformational state transitions with corresponding voltage-dependent rate constants labeled (e.g., $\alpha 13$ and $\beta 13$ are the rate constants for transitions from $C1$ to $O$ and $O$ to $C1$, respectively). (B) The full lidocaine-Na+ channel interaction model. The drug-free model from (A) is depicted in black. Red ($D^+$ prefix) and blue ($D$ prefix) states represent conformational states where charged and neutral drug is bound, respectively. Charged drug can only bind to non-inactivated states ($C3$, $C2$, $C1$, and $O$), while neutral drug can bind to any state. Drug binding and unbinding rates are state-dependent, as indicated by $kon$, $kcon$, $koff$, $kcoff$, etc. For clarity, blue (black) circles as opposed to arrows were used to indicate neutral drug binding (unbinding) to the fast inactivated states with a rate constant $ki_on$ ($ki_off$).*

The Moreno et al. model includes drug-channel interactions for both the neutral and charged forms of lidocaine. In Fig 1B, the interaction between charged lidocaine and $Na^+$ channels is shown by the red conformational state labels and transition arrows. Lidocaine's binding site is located in the interior of the $Na^+$ channel pore, which the charged form of lidocaine can only access from the intracellular side of a non-inactivated channel [9]. Hence, charged drug can only bind to and unbind from non-inactivated conformational states ($O$, $C1$, $C2$, and $C3$). Moreno et al. adopted the modulated receptor hypothesis [14]: When charged drug is bound to the $Na^+$ channel, the channel can undergo the same transitions as when no drug is bound but some transition rates are altered by drug binding.

The interaction of neutral lidocaine with $Na^+$ channels is shown by blue conformational state labels and transition arrows in Fig 1B. Again, Moreno et al. use the modulated receptor hypothesis. However, unlike charged lidocaine, neutral lidocaine can access the binding site when the channel is in either inactivated or non-inactivated states [9]. Therefore, Moreno et al. allow neutral lidocaine to bind and unbind to $Na^+$ channels in any conformational state, but the rate constants of drug binding are state-dependent.

Assuming mass action kinetics, the dynamics of the fraction of channels in each conformational state are given by a system of 7 differential equations for the no drug case and 23 differential equations for the full lidocaine-$Na^+$ channel interaction model (see S6 Appendix for equations).

## 2.2. Analysis of the Moreno et al. model shows that lidocaine preferentially binds to and stabilizes the inactivated state of $Na^+$ channels

Below, we examine the kinetics of the "detailed" Moreno et al. model and formulate a set of approximations about lidocaine's interaction with the $Na^+$ channel. We then use these assumptions to inform the structure of our low-dimensional lidocaine-$Na^+$ channel interaction model in Subsection 2.3.

- *Approximation 1: Charged lidocaine has no effect on $Na^+$ channel kinetics.*

Charged and neutral forms of lidocaine have similar concentrations at physiological pH. Using lidocaine's pKa of 7.6 [4, 9], Moreno et al. estimate that $\sim 60\%$ of lidocaine is positively charged and $\sim 40\%$ is neutral at physiological pH. On the other hand, the binding affinity of neutral lidocaine to inactivated channels is two orders of magnitude larger than that of charged lidocaine. Specifically, the dissociation constants, $K_d$, of neutral lidocaine binding to inactivated channels is $6.8\ \mu M$, whereas the $K_d$ for charged lidocaine binding to inactivated channels ranges from $188$ to $2590\ \mu M$ for physiological membrane potentials at $37°\mathrm{C}$ [4, 9-11]. Therefore, lidocaine bound to $Na^+$ channels will predominantly be in the neutral form (see S1 Appendix for further details), hence we will only include the effects of the neutral form of lidocaine in our low-dimensional model of the lidocaine-$Na^+$ channel interaction.

- *Approximation 2: Neutral lidocaine only binds to and unbinds from inactivated $Na^+$ channels.*

In the Moreno et al. model, neutral lidocaine can bind to both inactivated and non-inactivated conformational states. However, the $K_d$ of neutral lidocaine binding to inactivated channels ($6.8\ \mu M$) is two orders of magnitude smaller than that for binding to non-inactivated channels ($400\ \mu M$ and $1800\ \mu M$ for closed or open channels, respectively) [4, 9, 11]. Hence, we will assume that neutral lidocaine can only bind to inactivated channels in our low-dimensional model.

- *Approximation 3: Binding of neutral lidocaine locks $Na^+$ channels in the inactivated state until the drug unbinds.*

Simulations of the Moreno et al. model show that following lidocaine binding to $Na^+$ channels in the inactivated state, almost all channels remain in the inactivated state until drug unbinds (see S1 Appendix). This result arises from the transition rate constants of bound $Na^+$ channels in the Moreno et al. model, which are such that when drug binds to channels, the inactivated state is substantially stabilized. This stabilization of the inactivated state can be quantified by the "relative stability" of inactivated states, which we define as the ratio of the occupancy of non-inactivated states to the occupancy of inactivated states at steady state. Over the entire physiological range of $V$, the stability ratios for drug bound channels are always less than $0.1$, which implies that the binding of neutral lidocaine effectively locks the $Na^+$ channel in an inactivated conformation until drug unbinds (see S1 Appendix). Therefore, in our low-dimensional model, we assume that when neutral lidocaine binds to inactivated channels, drug bound channels cannot recover from inactivation until drug unbinds.

## 2.3. A low-dimensional model for lidocaine-$Na^+$ channel interaction

To construct a low-dimensional model of $Na^+$ conductance, we utilize a Hodgkin-Huxley formulation for drug-free $Na^+$ channel dynamics. The model has three activation gates and one inactivation gate; such that the fraction of open channels, $f_{open}$, is $f_{open} = m^3h$. The differential equations governing the dynamics of the fraction of activation and inactivation gates that are open ($m$ and $h$, respectively) are

$$\frac{dm}{dt} = \alpha_m(1 - m) - \beta_m m$$
$$\frac{dh}{dt} = \alpha_h(1 - h) - \beta_h h, \qquad (2)$$

where $\alpha_m$, $\beta_m$, $\alpha_h$ and $\beta_h$ are voltage-dependent rate constants of the form $c_1 e^{\frac{V}{c_2}}$ (parameters are provided in S7 Appendix).

We extend this low-dimensional $Na^+$ channel model to include the effects of lidocaine by implementing the assumptions from the preceding section. Specifically, we introduce an additional gating variable, $b$, which represents the fraction of channels bound to neutral lidocaine [15, 16] (by Approximation 1, charged lidocaine plays a negligible effect). Thus, in the presence of drug, the fraction of open channels is

$$f_{open} = m^3 h(1 - b).$$

To derive the equation governing the dynamics of $b$, note that neutral lidocaine can only bind to inactivated channels (by Approximation 2), so the binding rate is modulated by the fraction of channels that are inactivated and available for binding. Thus, the instantaneous drug binding rate is $k_{on}[D](1 - h)$. Then, as all drug bound channels are locked in the inactivated state (by Approximation 3), drug unbinding is not inhibited by non-inactivated channels “trapping” the drug, and the unbinding rate is simply $k_{off}$. Therefore, the dynamics of $b$ are governed by

$$\frac{db}{dt} = k_{on}[D](1-h)(1-b) - k_{off}b. \tag{3}$$

As in the Moreno et al. model, the drug binding rates are based on previously published values, $k_{on} = 250\ M^{-1}ms^{-1}$ and $k_{off} = 1.7\ \times 10^{-3}\ ms^{-1}$ [4, 9, 11]. By combining the low-dimensional $Na^+$ channel model in Equation (2) with the drug binding dynamics of Equation (3), the conductance of our low-dimensional lidocaine-$Na^+$ channel interaction model is $g_{Na} = \bar{g}_{Na}m^3h(1-b)$.

### 2.4. Our low-dimensional model is an order of magnitude lower in dimension than the Moreno et al. model

Our low-dimensional model of the drug-free $Na^+$ conductance has 2 variables and a set of 8 parameters that need to be optimized to fit experimental data, whereas the drug-free Moreno et al. model has 7 variables and 21 parameters [4]. The difference in complexity in the two models is even greater when drug-channel interactions are included in the models. The lidocaine component of our low-dimensional model consists of only 1 additional variable [15, 16] and 3 additional parameters for a total of 3 variables and 11 parameters. The lidocaine component of the Moreno et al. model has 16 additional variables and 19 additional parameters for a total of 23 variables and 40 parameters. In both lidocaine interaction models, values for drug binding rates are taken from the literature [9-11], and drug concentrations, $[D]$, are set by experimental protocols and pH calculations [4]. Therefore, no further parameter optimization is required to set the values of the 3 additional parameters of our low-dimensional model. On the other hand, in the Moreno et al. model, 10 further parameters describing drug-dependent state transition rates were required to be optimized to data. Altogether, our lidocaine-$Na^+$ channel interaction model is an order of magnitude lower in dimension than the Moreno et al. model (3 compared to 23 variables) and has nearly four times fewer free parameters (8 compared to 31).

## 3. Results

### 3.1. Both the Moreno et al. and low-dimensional models reproduce experimental voltage-clamp data

The validity of the Moreno et al. model and our low-dimensional model are assessed by their abilities to reproduce previously published voltage-clamp data [9, 10, 17-20] that characterize the kinetics of the $Na^+$ conductance in the presence and absence of lidocaine. In this section, we fit our low-dimensional model to the same experimental data that Moreno et al. used to fit their model and then compare the quality of fits for the two models. The fits to individual experimental voltage-clamp protocols are examined for drug-free conditions (Subsection 3.1.1) and in the presence of lidocaine (Subsection 3.1.2). We find that the low-dimensional model performs similarly well to the Moreno et al. model in recapitulating the data, despite having fewer free parameters for fitting to data.

**3.1.1. Drug-free models.** To appropriately compare the Moreno et al. model and our low-dimensional model of the lidocaine-$Na^+$ channel interaction, the baseline $Na^+$ conductance models need to exhibit similar kinetics. Therefore, we fit our drug-free Hodgkin-Huxley ($m^3h$) $Na^+$ conductance model, Equation $(2)$, to voltage-clamp experimental data that Moreno et al. used to fit their model. Specifically, we fit the low-dimensional model to the steady state availability [17], steady state activation [18], and time to half inactivation data [18, 20] (Fig 2A, B, and C) that was used to constrain the Moreno et al. model. To constrain the activation rates, the low-dimensional model is also fit to activation rate data (Fig 2D) [19]. Full experimental protocol details are provided in the S2 Appendix.

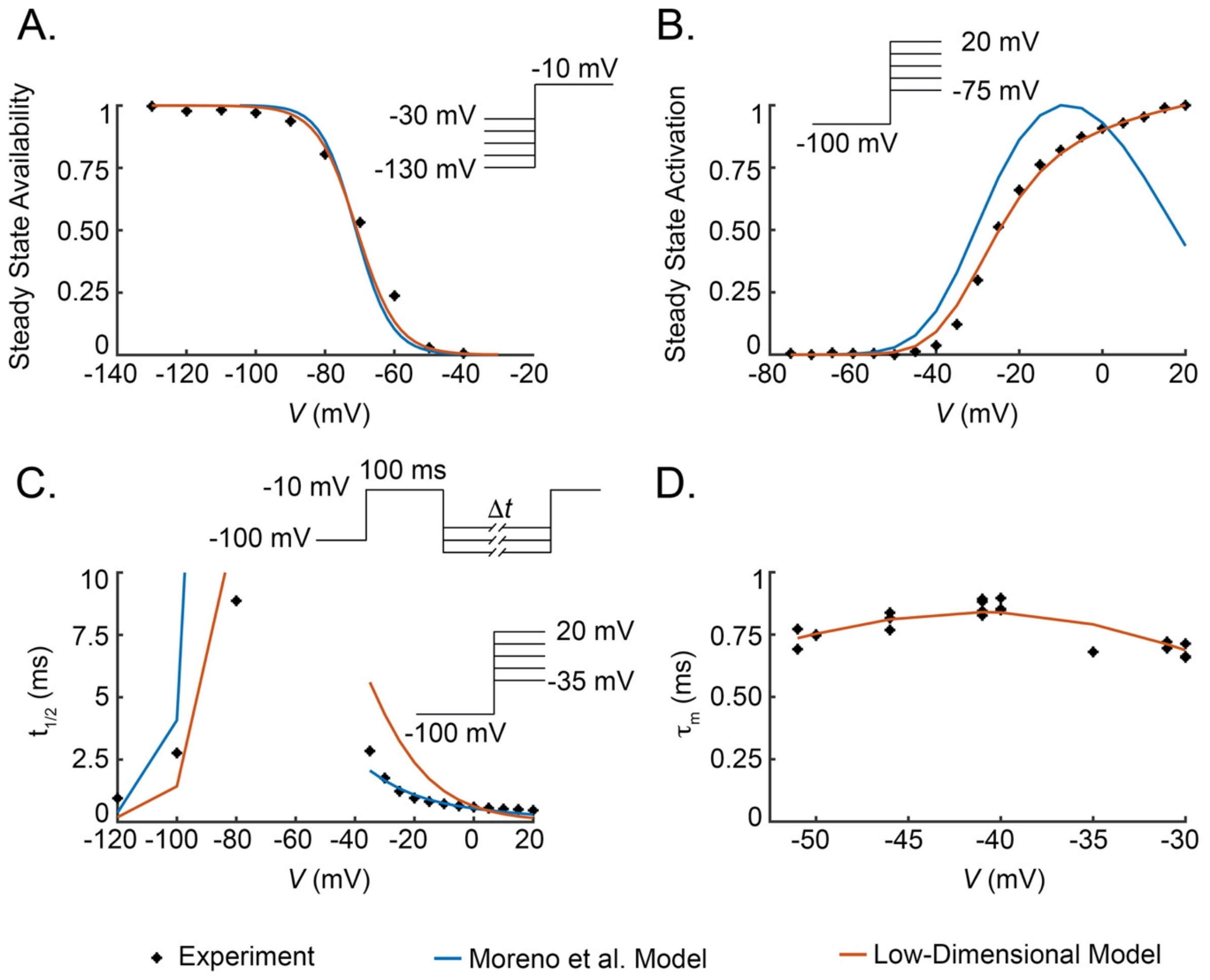

***Fig 2. Drug-free data.*** *The low-dimensional Na⁺ conductance model is fit to experimental data from steady state availability (A; sum squared errors (SSE) of* $0.038$ *and* $0.035$ *for the Moreno et al. and low-dimensional models, respectively), steady state activation (B; SSE of* $0.042$ *and* $7.0 \times 10^{-4}$ *for the Moreno et al. and low-dimensional models, respectively), time to half inactivation (C; SSE of* $197$ *and* $17$ *for the Moreno et al. and low-dimensional models, respectively), and time constant of activation (D; SSE of* $1.4 \times 10^{-3}$ *for the low-dimensional model) voltage-clamp experiments [17-19]. In all subfigures, black asterisks (*) indicate experimental data points, blue curves represent output from the Moreno et al. model [4], orange curves represent output from the low-dimensional Na⁺ conductance model.* $t_{1/2}$ *at* $-80$ $mV$ *in the Moreno et al. model is not indicated in the figure as it is substantially larger* ($48.8$ $ms$) *than the other* $t_{1/2}$ *values. For the low-dimensional model in (D),* $\tau_m = \frac{1}{\alpha_m + \beta_m}$. $SSE = \frac{1}{n}\sum_{i=1}^{n}(y_i - x_i)^2$, $n$ *is total number of data points,* $\{y_i\}$ *are the model output, and* $\{x_i\}$ *are the experimental data values.*

Fig 2 illustrates that, despite our low-dimensional model having only 8 free parameters and the Moreno et al. model having 21 free parameters, both models are able to replicate the experimental data. In particular, the Moreno et al. model and our low-dimensional model exhibit very similar output for steady state availability (Fig 2A) and time constant of inactivation (Fig 2C). In the steady state activation protocol (Fig 2B), our low-dimensional model outperforms the Moreno et al. model. Finally, the activation time constants of the low-dimensional model closely match experimentally determined activation rate constants (Fig 2D).

**3.1.2. Lidocaine-$Na^+$ channel models.** Having established the parameterization of the drug-free models, we now examine the ability of the lidocaine components of the models to reproduce previously published voltage-clamp data [9, 10, 17] that characterize the effects of lidocaine. As stated in Subsection 2.4, for both models, the drug binding and unbinding rates were set to previously published values [9-11]. We stress at the onset that no further parameters in our low-dimensional model were fit to this data, whereas 10 additional parameters were used to fit the Moreno et al. model to the data. Therefore, the results that follow show the fits of the Moreno et al. model but strictly serve only as validation for our low-dimensional drug-channel interaction model.

We consider five voltage-clamp data sets, including steady state availability, tonic block (i.e., the steady state fraction of channels bound to drug), dose-dependence of use-dependent block, recovery from use-dependent block, and frequency dependence of block (details of the experimental protocols are provided in S2 Appendix) [9, 10, 17]. Fig 3 displays the voltage-clamp data in the absence and presence of lidocaine (asterisks and circles), and the corresponding output from the Moreno et al. model (blue curves) and our low-dimensional model (orange curves).

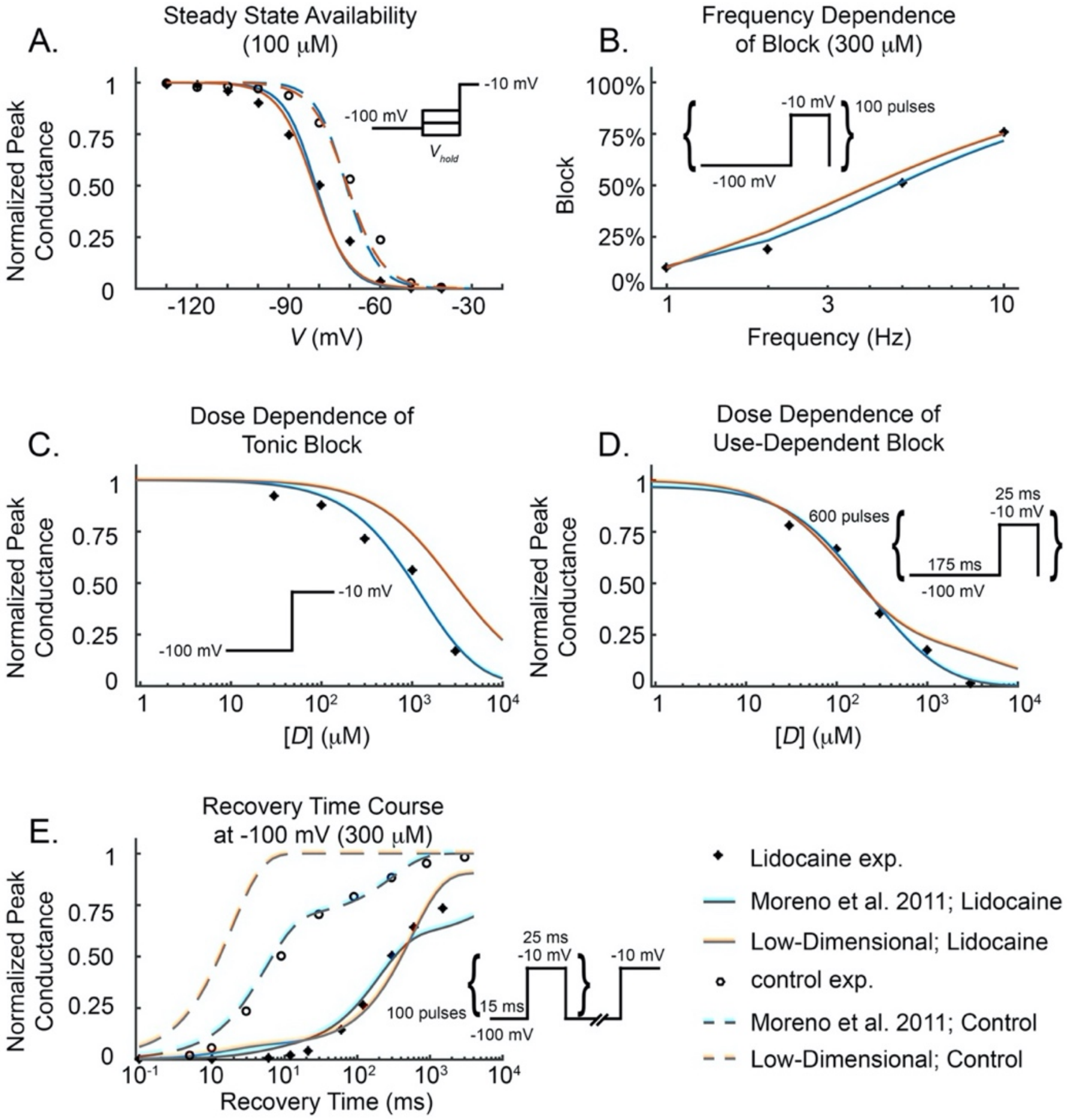

***Fig 3. Lidocaine voltage-clamp data.*** *Experimental voltage-clamp data (asterisks), Moreno et al. 2011 model fit (solid blue curves), low-dimensional model output (solid orange curves) for lidocaine effects on the Na⁺ conductance.* $V$*-clamp data for control, i.e. no drug, experiments (circles), Moreno et al. (blue dashed curves), and low-dimensional model with no drug (orange dashed curves) are included for appropriate protocols. (A) Steady state availability with SSE of* $3.6 \times 10^{-3}$ *and* $2.9 \times 10^{-3}$ *for the Moreno et al. and low-dimensional lidocaine models, respectively. (B) Frequency dependence of block with SSE of* $8.8 \times 10^{-4}$ *and* $1.7 \times 10^{-3}$ *for the Moreno et al. and low-dimensional lidocaine models, respectively. (C) Tonic block with SSE of* $3.1 \times 10^{-3}$ *and* $0.033$ *for the Moreno et al. and low-dimensional lidocaine models, respectively. (D) Dose-dependence of use-dependent block with SSE of* $1.3 \times 10^{-3}$ *and* $1.1 \times 10^{-2}$ *for the Moreno et al. and low-dimensional lidocaine models, respectively. (E) Recovery from use-dependent block with SSE of* $2.3 \times 10^{-3}$ *and* $5.0 \times 10^{-3}$ *for the Moreno et al. and low-dimensional lidocaine models, respectively.* $SSE = \frac{1}{n}\sum_{i=1}^{n}(y_i - x_i)^2$, $n$ *is total number of data points,* $\{y_i\}$ *are the model output, and* $\{x_i\}$ *are the experimental data values [9, 10, 17].*

Steady state availability predicted by our low-dimensional model for lidocaine-$Na^+$ channel interaction (Fig 3A; orange curve) agrees well with the experimental data (asterisks) [9], as does the fit of the Moreno et al. model (blue curve). The frequency dependence of block protocol (Fig 3B) [10], is also closely replicated by the low-dimensional and Moreno et al. models. Our low-dimensional model replicates the dose-dependence of tonic block and use-dependent block data (Fig 3C and D, respectively) [10] at the clinically relevant range of $5$ to $20$ $\mu M$ [21], but slightly under predicts the fraction of blocked channels at drug concentrations much higher than the clinically relevant range. (The disparity between the low-dimensional model and tonic block data above the clinically relevant range is explained in S3 Appendix.) The Moreno et al. model performs slightly better at replicating dose-dependent block at the higher concentrations examined experimentally. Despite the recovery from inactivation of the low-dimensional model being too quick in the drug-free case (Fig 3E, dashed orange curve), the recovery from drug binding at $-100$ $mV$ time course of the low-dimensional model (solid orange curve) performs similarly well to the Moreno et al. model (solid blue curve) in replicating the data.

### 3.2. Functional effects of lidocaine

The previous subsection demonstrated that both the low-dimensional model and the Moreno et al. model are able to replicate $Na^+$ conductance voltage-clamp data both in the presence and absence of lidocaine. However, the primary role of mathematical models of drug-ion channel interactions is not simply to fit/predict voltage-clamp experimental results; it is to identify the mechanisms by which drugs affect dynamics at the cellular and tissue levels (e.g., peak upstroke velocity and conduction velocity), and therefore alter the propensity for arrhythmias [2, 5]. First, we run simulations to systematically examine and compare the rate-dependent effects of lidocaine on peak upstroke velocity and conduction velocity in the Moreno et al. and low-dimensional models. Then, we utilize the relatively simple structure of the low-dimensional model to uncover the mechanisms underlying lidocaine's rate-dependent effects.

**3.2.1. Low-dimensional model and Moreno et al. model predict similar functional effects of lidocaine on upstroke velocity and conduction velocity.** In order to examine the rate-dependent effects of lidocaine on peak upstroke velocity and conduction velocity of the cardiac action potential, we incorporate both lidocaine-$Na^+$ channel interaction models into the ten Tusscher et al. model for human ventricular myocytes [22, 23] (see S4 Appendix for details of implementation of $Na^+$ conductance models into the ten Tusscher et al. model). Peak upstroke velocities (and conduction velocities) in cases with drug present are normalized by the drug-free peak upstroke velocity (conduction velocity) at the same basic cycle length (BCL). (Non-normalized peak upstroke velocities and conduction velocities are provided in S4 Appendix.)

Fig 4A shows normalized peak upstroke velocities for BCLs ranging from $300$ $ms$ to $1000$ $ms$ (i.e., the full "physiological range") for the Moreno et al. model (blue curves) and low-dimensional model (orange curves) with drug concentrations of $5$ $\mu M$ (solid curves) and $20$ $\mu M$ (dashed curves), which correspond to low and high clinical plasma concentrations [21]. Both the low-dimensional and Moreno et al. models predict larger decreases in peak upstroke velocity due to lidocaine block as BCL decreases. At a drug concentration of $20\ \mu M$, both models predict approximately a $10\%$ reduction in peak upstroke velocity at a BCL of $1000\ ms$ ($7.5\%$ and $13.5\%$ reductions in the Moreno et al. and low-dimensional models, respectively), whereas upstroke velocity is decreased by approximately $30\%$ at BCL of $300\ ms$ ($29\%$ and $35\%$ reductions in the Moreno et al. and low-dimensional models, respectively).

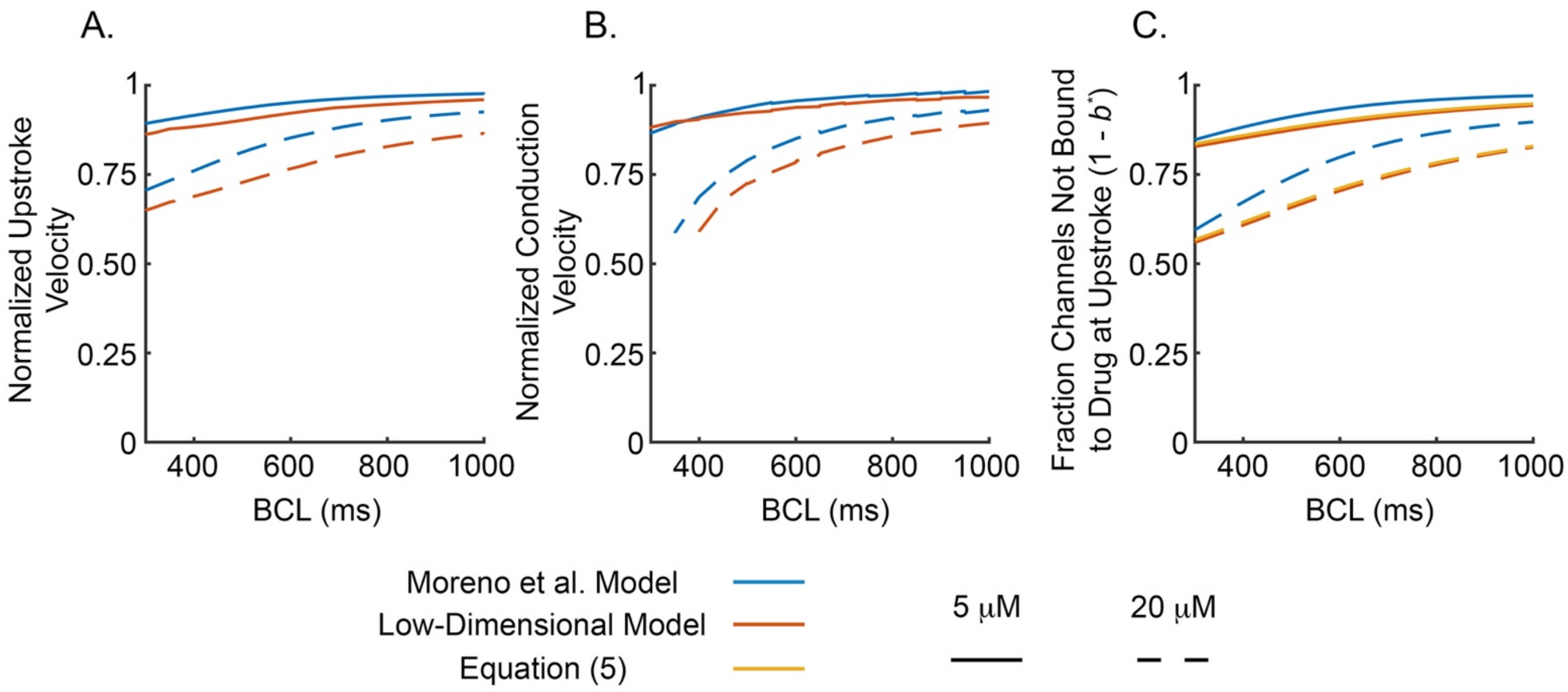

***Fig 4. Rate-Dependent Effects of Lidocaine.*** *Normalized peak upstroke velocity (A), conduction velocity (B), and fraction of channels not bound to drug during the upstroke (C) plotted against BCL for the ten Tusscher et al. human ventricular myocyte model [22, 23] with the Moreno et al. model (blue curves) or low-dimensional model (orange curves) of the* $Na^+$ *conductance. Peak upstroke and conduction velocities for* $5\ \mu M$ *(solid curves) and* $20\ \mu M$ *(dashed curves) concentrations of lidocaine are normalized by peak upstroke and conduction velocities at the same BCL in the corresponding drug-free model. In C, predictions from an analytically derived expression for fraction of channels bound to drug during the upstroke (*$b^*$*) in our low-dimensional model are also plotted (yellow curves; see equation (5)) and were shifted up slightly to make them visible as they overlap the output of the ten Tusscher et al. model with our low-dimensional* $Na^+$ *conductance model.*

Fig 4B displays normalized conduction velocities for lidocaine concentrations of $5\ \mu M$ (solid curves) or $20\ \mu M$ (dashed curves) as a function of BCL. Note that the curves for $20\ \mu M$ lidocaine terminate at BCLs of $350$ and $400$ $ms$ for the Moreno et al. (blue curves) and low-dimensional (orange) models, respectively, because pacing with shorter BCLs did not elicit propagating waves for each stimulus. As with upstroke velocity, both models predict more prominent effects of lidocaine block as BCL decreases, i.e., there is a greater reduction in conduction velocity at lower BCL. Specifically, at a lidocaine concentration of $20$ $\mu M$, both models predict approximately a $10\%$ reduction in normalized conduction velocity for a BCL of $1000$ $ms$ ($7\%$ and $11\%$ for the Moreno et al. and low-dimensional models, respectively), and this reduction increases to approximately $35\%$ as BCL decreases to $400$ $ms$ ($31\%$ and $41\%$ for the Moreno et al. and low-dimensional models, respectively).

The available $Na^+$ conductance during the upstroke, and hence the fraction of channels bound to drug during the upstroke ($b^*$), is considered to be one of the primary determinants of peak upstroke velocity and conduction velocity [2]. The fraction of channels *not* bound to drug during the upstroke ($1 - b^*$) as a function of BCL in the ten Tusscher et al. model with either the Moreno et al. (blue curves) or our low-dimensional (orange curves) $Na^+$ conductance models are plotted in Fig 4C. $1 - b^*$ has the same rate-dependence as normalized peak upstroke and conduction velocity, verifying the strong causal link between these quantities and the level of block of $Na^+$ channels by lidocaine.

**3.2.2. Rate-dependent effects of lidocaine arise from voltage-dependence of Na⁺ channel inactivation: Insight from analysis of low-dimensional model.** In this section, we analyze our low-dimensional model, and by capitalizing on its relative simplicity, we elucidate the mechanisms underlying the rate-dependent effects of lidocaine on upstroke velocity and conduction velocity. Specifically, we derive an expression for the fraction of $Na^+$ channels bound to drug during the upstroke ($b^*$), similar to that derived by Starmer et al. [24] and Weirich and Antoni [25]. This expression captures the parametric dependence on BCL, action potential duration restitution properties, diastolic potential, channel inactivation kinetics, and drug concentration and binding rates.

We first note that, when the ten Tusscher et al. model with our low-dimensional $Na^+$ conductance model (from now on referred to as "our modified ten Tusscher et al. model") is paced at a constant BCL, the membrane potential ($V$) is approximately -85 $mV$ during the diastolic interval (DI), and $V$ is approximately 20 $mV$ for the duration of the action potential (APD) (Fig 5A; blue curve). Therefore, we approximate the time course of $V$ by a square wave alternating between -85 $mV$ and 20 $mV$ (Fig 5A; orange curve). Furthermore, throughout the DI and AP, the fraction of non-inactivated $Na^+$ channels, $h$, is approximately at the steady state values of $h_\infty(-85\ mV) \approx 0.9$ and $h_\infty(20\ mV) \approx 0$, respectively. Consequently, we approximate the time course of $h$ as a square wave alternating between $h_\infty(-85\ mV)$ and $h_\infty(20\ mV)$. For a given BCL, we take the APD and DI of our square wave approximations to be given by the $APD_{90}$ restitution curve of our modified ten Tusscher et al. model (i.e., APD is a function of BCL as displayed by the blue curve in Fig 6A, and $DI = BCL - APD$), noting that the effect of lidocaine on the restitution curve is negligible other than shifting the minimum BCL at which an action potential is induced following each stimulus.

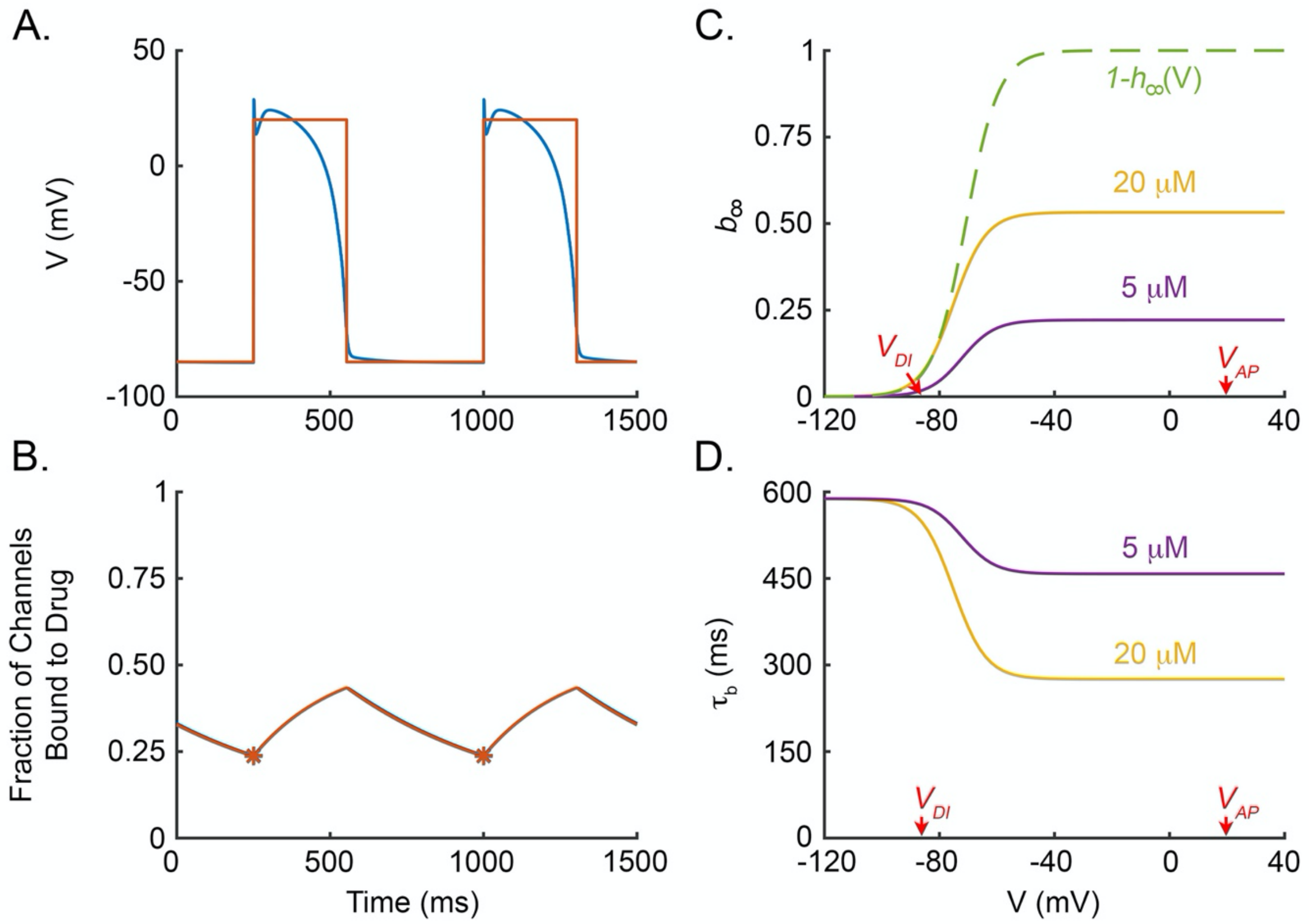

***Fig 5. Mechanism of lidocaine rate-dependent effects.*** *(A) Simulated action potentials of our modified ten Tusscher et al. model paced at a BCL of $750\ ms$ with $20\ \mu M$ lidocaine (blue) and a square wave approximation of the action potential (orange) that alternates between -$85\ mV$ and $20\ mV$. (B) Dynamics of the fraction of channels bound to drug, $b$, with $20\ \mu M$ lidocaine present in our modified ten Tusscher et al. model (blue) and low-dimensional $Na^+$ conductance model stimulated by the square wave approximation (orange). (C) Voltage dependence of $1 - h_\infty$ (dashed green) and $b_\infty$ for $20\ \mu M$ (yellow) and $5\ \mu M$ (purple) lidocaine. (D) Voltage dependence of $\tau_b$ for $20\ \mu M$ (yellow) and $5\ \mu M$ (purple) lidocaine.*

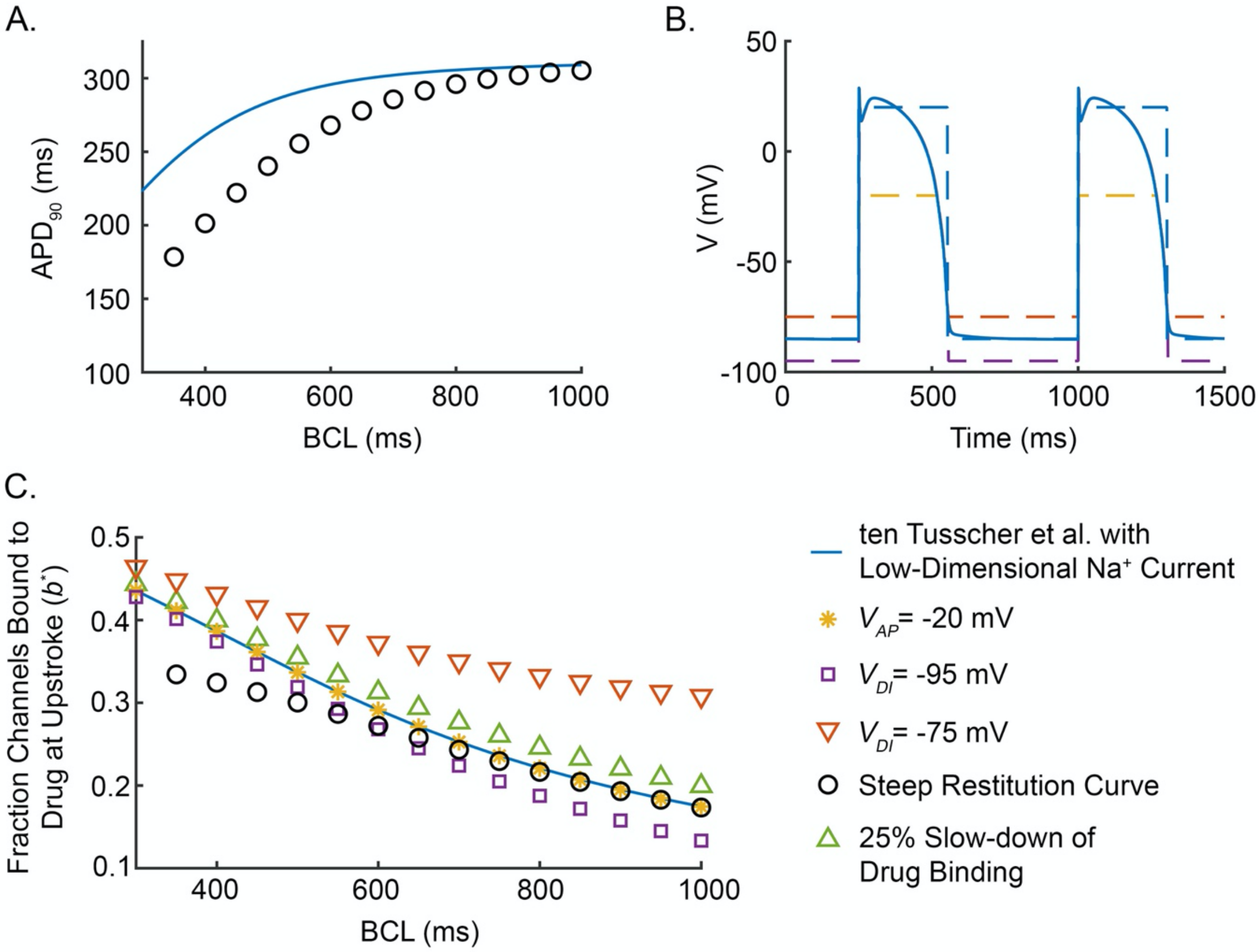

***Fig 6. Cardiac electrophysiological properties and lidocaine binding.*** *(A) $APD_{90}$ restitution curve for our modified ten Tusscher et al. model with $20\ \mu M$ of lidocaine (blue curve) and a steeper, hypothetical restitution curve (black circles). (B) Time course of transmembrane potential for our modified ten Tusscher et al. model paced at a BCL of $750\ ms$ and the square wave approximation of $V$ (dashed blue line) as well as hypothetical alterations: $V_{AP}$ decreased to -$20\ mV$ (dashed yellow curve), $V_{DI}$ decreased to -$95\ mV$ (dashed purple curve), and $V_{DI}$ increased to -$75\ mV$ (dashed orange curve). (C) Fraction of channels bound to lidocaine during the upstroke ($b^*(BCL)$) in the presence of $20\ \mu M$ lidocaine, as given by Equation $(5)$, for our modified ten Tusscher et al. model and the various shifts in cardiac electrophysiological characteristics in (A) and (B).*

By exploiting the piecewise constant nature of the square wave approximations of $V$ and $h$, we are able to solve Equation $(3)$ to obtain formulae describing the fluctuations of the fraction of $Na^+$ channels bound to drug for arbitrary BCL,

$$
b(t) = \begin{cases} b_\infty(V_{AP}) - \big(b_\infty(V_{AP}) - b(t_k)\big)\, e^{-\frac{t-t_k}{\tau_b(V_{AP})}}, & t_k \le t < t_k + APD \\ b_\infty(V_{DI}) - \big(b_\infty(V_{DI}) - b(t_k + APD)\big)\, e^{-\frac{t-(t_k+APD)}{\tau_b(V_{DI})}}, & t_k + APD \le t < t_{k+1} \end{cases}, \qquad (4)
$$

where $t_k$ is the time of the k$^{th}$ upstroke, $b(t_k)$ is the fraction of channels bound to lidocaine at the time of the upstroke, and

$$
\tau_b(V) = \frac{1}{\big(1 - h_\infty(V)\big)[D]k_{on} + k_{off}}, \qquad b_\infty(V) = \frac{\big(1 - h_\infty(V)\big)[D]k_{on}}{\big(1 - h_\infty(V)\big)[D]k_{on} + k_{off}}.
$$

Fig 5B shows that the dynamics of the fraction of channels bound to drug, $b(t)$, given by the approximation in Equation (4) and direct simulations of our modified ten Tusscher et al. model (orange and blue curves, respectively) are in such close agreement that the curves are nearly indistinguishable.

The steady state fraction of channels bound to drug during the upstroke ($b^*$) can be found by setting the fraction of channels bound to drug during the upstroke at time $t_k$ equal to the fraction of channels bound during the subsequent upstroke at time $t_{k+1}$,

$$
b(t_k) = b(t_{k+1}) = b^*.
$$

Solving this equation yields

$$b^* = \left[\frac{1-D}{1-AD}\right] b_\infty(V_{DI}) + \left[\frac{(1-A)D}{1-AD}\right] b_\infty(V_{AP}), \tag{5}$$

where $A = e^{-\frac{APD}{\tau_b(V_{AP})}}$ and $D = e^{-\frac{DI}{\tau_b(V_{DI})}}$. Figure 4C plots the rate-dependence of $1 - b^*$ as predicted by the algebraic expression in Equation (5) (yellow curves) along with the results from direct simulations of our modified ten Tusscher et al. model (orange curves). These curves are indistinguishable on the scale of the figure, indicating that Equation (5) provides an excellent approximation for the level of drug binding during periodic pacing.

Equation (5) implies that, under constant pacing, the fraction of channels bound to drug during the upstroke ($b^*$) is a weighted sum of the steady state fraction of channels bound to drug at the plateau potential ($b_\infty(V_{AP})$) and the steady state fraction of channels bound to drug at the diastolic potential ($b_\infty(V_{DI})$) (Fig 5C). The weight factors are dependent on BCL through APD and DI as dictated by the restitution curve (blue curve in Fig 6A), as well as the effective time constants of drug binding at the plateau potential ($\tau_b(V_{AP})$) and the diastolic potential ($\tau_b(V_{DI})$) (Fig 5D). Thus, Equation (5) provides an efficient way to examine the dependence of drug binding on heart rate and cellular dynamics (i.e., through restitution properties, and diastolic and plateau potentials), as well as drug concentration and binding kinetics. Furthermore, analysis of Equation (5) can help identify the mechanisms underlying the rate-dependencies.

- *Steeper restitution curve decreases rate-dependence of lidocaine binding.*

Fig 6A displays the restitution curve of our modified ten Tusscher et al. model (blue curve) and a hypothetical, steeper restitution (black circles), which we use to illustrate the effect of restitution properties on the rate-dependence of lidocaine block. Fig 6C plots the corresponding rate-dependence of the fraction of $Na^+$ channels blocked by drug during the upstroke of the action potential ($b^*$) as predicted by Equation (5) in the presence of $20\ \mu M$ of lidocaine. At high BCL, $b^*$ increases as BCL decreases for both the default and the steeper restitution curves. For lower BCL, where APDs in the steeper restitution curve are substantially shorter than in our modified ten Tusscher et al. model, $b^*$ continues to increase as BCL decreases, but the steeper restitution curve yields a lower level of block ($b^*$) and a lower degree of rate-dependence than the default case.

The lower level of block in the case of the steeper restitution curve is easily understood. Note that the voltage dependence of $b_\infty(V)$, which is inherited from $h_\infty(V)$, indicates that lidocaine tends to bind during the AP and unbind during the diastolic interval (Fig 5C). The steeper restitution curve is associated with a smaller APD and a correspondingly larger DI for a given BCL. Therefore, there is less time for drug binding during the action potential and more time for unbinding during the DI, leading to an overall decrease in the level of bound drug. This effect is embedded in Equation (5) as a decreased weight of $b_\infty(V_{AP})$ and increased weight of $b_\infty(V_{DI})$ through their dependence on APD and DI.

The increased level of block ($b^*$) with decreased BCL at high BCL can be explained in a similar manner. The restitution curve is flat for large BCL, and therefore APD remains roughly constant while DI accounts for any change in BCL. This implies that, as BCL decreases, there is less time for drug unbinding during the DI, which leads to an overall increase in the level of bound drug.

It is not immediately clear how the level of block ($b^*$) should change with decreases in BCL at lower BCL (i.e., the portion of the restitution curves where both DI and APD change substantially with BCL). However, insight can be obtained from the sensitivity of $b^*$with respect to BCL according to Equation (5),

$$\frac{\partial b^*}{\partial BCL} = \frac{b_\infty(V_{AP}) - b_\infty(V_{DI})}{(1-AD)^2}\left(\frac{(1-D)AD}{\tau_b(V_{AP})}\frac{df(BCL)}{dBCL} - \frac{(1-A)D}{\tau_b(V_{DI})}\left(1 - \frac{df(BCL)}{dBCL}\right)\right), \quad (6)$$

where $APD = f(BCL)$ is the restitution curve (see S5 Appendix for derivation). First note that when the restitution curve is flat (i.e., $\frac{df(BCL)}{dBCL} \approx 0$), Equation (6) confirms that $b^*$increases with decreasing BCL (as described above). However, as the restitution curve steepens, (i.e., $\frac{df(BCL)}{dBCL} > 0$), Equation (6) indicates that degree of rate-dependence will decrease. In fact, Equation (6) suggests that reverse rate-dependence is possible if

$$\frac{(1-D)AD}{\tau_b(V_{AP})}\frac{df(BCL)}{dBCL} > \frac{(1-A)D}{\tau_b(V_{DI})}\left(1 - \frac{df(BCL)}{dBCL}\right).$$

In other words, there is a critical slope of the restitution curve above which reverse rate-dependent drug binding occurs,

$$\frac{df(BCL)}{dBCL} = \left(1 + \frac{\tau_b(V_{DI})}{\tau_b(V_{AP})}\frac{(1-D)A}{(1-A)}\right)^{-1}.$$

For the drug binding rates of lidocaine, we find that reverse rate-dependence would only occur for non-physiological restitution curves. However, reverse rate-dependence could occur under physiological conditions for state-dependent $Na^+$ channel blockers with similar binding pathways but slower unbinding rates.

- *Lidocaine binding is insensitive to changes in AP amplitude and plateau potential.*

To examine the effect of the AP plateau potential ( $V_{AP}$) on the rate-dependent binding of lidocaine, we shift $V_{AP}$ from $20\ mV$ in our default model (Fig 6B; dashed blue curve) to a hypothetical value of -20 $mV$ (dashed yellow curve). Fig 6C shows that the fraction of channels bound to drug during the upstroke ($b^*$) for $V_{AP} = -20\ mV$ (Fig 6C; yellow asterisks) overlays the values for $V_{AP} = 20\ mV$ (blue curve).

Equation (5) reveals that the insensitivity of $b^*$ on $V_{AP}$ is inherited entirely through its dependence on $b_\infty(V_{AP})$ and $\tau_b(V_{AP})$, which in turn inherit their dependence on $V_{AP}$ through the steady state inactivation curve $1 - h_\infty(V_{AP})$. (The quantitative details of the local insensitivity of $b^*$ on $V_{AP}$ are given by the derivative $\frac{\partial b^*}{\partial V_{AP}}$; see S5 Appendix.) The dashed green curve in Fig 5C demonstrates that steady state inactivation is saturated at potentials above -55 $mV$, and thus, $b^*$ is insensitive to variations in $V_{AP}$. In fact, the insensitivity of $b^*$ to variations in $V_{AP}$ implies that, while lidocaine block in our modified ten Tusscher et al. model depends on the duration of the AP, it does not depend on the detailed shape of the AP.

- *Increases in the diastolic potential promotes lidocaine binding, yet binding is insensitive to decreases in diastolic potential.*

The effects of decreasing and increasing the diastolic potential $V_{DI}$ by $10\ mV$ on the fraction channels bound to drug during the upstroke ($b^*$) are illustrated in Fig 6C (purple squares and orange inverted triangles, respectively). Decreasing $V_{DI}$ decreases $b^*$ marginally at low BCL and moderately at high BCL. On the other hand, increasing $V_{DI}$ can substantially increase $b^*$, especially at longer BCL. As was the case for $V_{AP}$, the sensitivity of $b^*$ to $V_{DI}$ is inherited from the steady state inactivation curve (Fig 5C). In our modified ten Tusscher et al. model, $V_{DI} \approx -85\ mV$, which is just below the threshold of the steady state inactivation curve. Below $V_{DI} = -85\ mV$, the steady state inactivation curve is flat with $1 - h_{\infty}(V) \approx 0$, causing $b^*$ to be insensitive to decreases in $V_{DI}$; whereas above $V_{DI} = -85\ mV$, $1 - h_{\infty}(V)$ steepens drastically, leading $b^*$ to be highly sensitive to increases in $V_{DI}$. (The quantitative details of the local sensitivity of $b^*$ on $V_{DI}$ are given by the derivative $\frac{\partial b^*}{\partial V_{DI}}$; see S5 Appendix.)

- *Lidocaine binding is insensitive to minor changes in drug binding rate.*

Fig 6C also displays the predicted effect that decreasing lidocaine binding and unbinding rates by 25% would have on the fraction of channels bound to lidocaine during the upstroke (green triangles). The predictions for $b^*$ are only slightly higher than those obtained when we used the published estimates of $k_{on}$ and $k_{off}$ [4, 9, 11], indicating that $b^*$ is locally insensitive to changes in the time constant of lidocaine binding kinetics. Note that equation (5) indicates that scaling $k_{on}$ and $k_{off}$ by a factor of $\sigma$ is equivalent to scaling APD and DI by $\sigma$, and thus its effect is the same as rescaling BCL with an adjusted restitution curve, such that

$$APD = \sigma^{-1} f(\sigma\ BCL)$$

(e.g., decreasing lidocaine binding and unbinding rates by 25% is equivalent to decreasing APD and BCL by 25%).

## 4. Discussion

In this study, we construct and analyze a novel low-dimensional model for lidocaine-$Na^+$ channel interaction. The structure of our model is based on (1) the mathematical framework proposed by Starmer et al. [15, 16, 24] for modeling state-dependent drug binding and (2) the key features of lidocaine-$Na^+$ channel interactions that we identify through an analysis of a high-dimensional Markov model by Moreno et al [4]. Our low-dimensional model consists of a two-variable Hodgkin-Huxley-type $Na^+$ conductance model and an additional variable for the fraction of channels bound to drug. Despite its low-dimensionality our model fits data from an extensive set of voltage-clamp experiments to a similar degree as the Moreno et al. model. Furthermore, similar effects of lidocaine on action potential upstroke velocity and conduction velocity are predicted when either our model or the Moreno et al. model is incorporated into the ten Tusscher et al. model for human ventricular cells.

The results from previous voltage-clamp experiments and computer simulation studies have suggested that lidocaine preferentially binds to and stabilizes $Na^+$ channels in the inactive state [11, 26-28], while other studies have suggested that lidocaine binds to $Na^+$ channel in other conformational states as well [9, 11]. Indeed, the Moreno et al. model allows lidocaine to bind to $Na^+$ channels in any state. However, our analysis of the Moreno et al. model demonstrates that the primary interaction of lidocaine with the $Na^+$ channel is stabilization of the inactivation gate. Specifically, our analysis of the state transition and binding rate constants of the Moreno et al. model identifies that (1) The vast majority of lidocaine bound to $Na^+$ channels is the neutral form of lidocaine; (2) Neutral lidocaine binds and unbinds almost exclusively to inactivated $Na^+$ channels; and (3) Upon binding to $Na^+$ channels, lidocaine effectively immobilizes the inactivation gate, such that channels cannot recover from inactivation before lidocaine unbinds. Our low-dimensional model is derived based on these three key properties of lidocaine-$Na^+$ channel interactions. The close agreement between extensive voltage-clamp data and our low-dimensional model supports the hypothesis that lidocaine binding to non-inactivated channels has a negligible contribution to lidocaine's overall effects. While the effects of charged lidocaine and binding to non-inactivated channels could be included in our model, it would come at the expense of higher dimensionality [29] and would likely only marginally improve the fits to the voltage-clamp data.

By exploiting the low-dimensionality of our model, we derive an algebraic expression for the fraction of channels bound to drug during periodic pacing ($b^*$). Previously, Starmer et al. [24] and Weirich and Antoni [25, 30] proposed similar expressions. We extend this previous work in two ways. First, we validate the key assumption that the drug binding rates during the action potentials and diastolic intervals are well approximated as constant values. Second, we use the algebraic expression for $b^*$ to explicitly explore the dependence of rate-dependent block on electrophysiological properties of cardiac cells (i.e., APD, DI, $V_{AP}$, and $V_{DI}$). In particular, we show that, while the level of lidocaine binding is highly dependent on action potential duration, lidocaine binding is unaffected by changes in the shape of the AP waveform (e.g., changes in plateau potential). The insensitivity of drug binding to the AP waveform results from steady state inactivation being saturated above -55 $mV$, which leads to an approximately constant binding rate of lidocaine for membrane potentials above -55 $mV$. On the other hand, because steady state inactivation, and hence lidocaine binding rate, shifts rapidly between -85 $mV$ and -75 $mV$, the level of lidocaine binding is highly sensitive to increases in diastolic potential ( $V_{DI}$).

To explore the functional effect of lidocaine on action potential upstroke velocity and conduction velocity, we incorporate our low dimensional model and the Moreno et al. model for drug-channel interactions into the ten Tusscher et al. model of human ventricular cells. However, it is important to note that quantitative predictions can be highly dependent on the details of the electrophysiological models utilized [31-33]. The algebraic expression for the fraction of channels bound to drug during the action potential upstroke provides a model-independent method for assessing the influence of electrophysiological properties of cardiac cells (i.e., APD, DI, $V_{AP}$, and $V_{DI}$) on lidocaine binding. Hence, our qualitative observations on how lidocaine binding is affected by changes in the AP waveform or restitution properties are robust to variation in the details of the electrophysiological models.

Lidocaine is considered to have a strong cardiac safety profile [5]. While we characterize the mechanisms underlying the rate-dependent block of lidocaine in the present study, the hallmarks of a safe $Na^+$ channel blocker cannot be determined from studying lidocaine alone. Rather, it requires a detailed comparison of the mechanisms underlying lidocaine's rate-dependent effects to those of less safe $Na^+$ channel blockers, such as flecainide [3]. To this end, low-dimensional models for $Na^+$ channel blockers with low safety profiles that include only primary binding pathways could be developed using the methods that we have employed here. The knowledge gained through analysis and comparison of such models could be invaluable in understanding the drug properties that characterize strong or weak safety profiles and thus greatly enhance our ability to develop new antiarrhythmic $Na^+$ channel blockers that are both safe and effective.

## Supporting Information

**S1 Appendix. Justification of lidocaine-$Na^+$ channel interaction approximations.**

**S2 Appendix. Experimental voltage-clamp protocols.**

**S3 Appendix. Low-dimensional model under predicts tonic block at -100 $mV$.**

**S4 Appendix. Peak upstroke and conduction velocity in modified ten Tusscher et al. models.**

**S5 Appendix. Dependence of lidocaine binding on physiological properties.**

**S6 Appendix. Moreno et al. model.**

**S7 Appendix. Equations of the low-dimensional lidocaine-$Na^+$ channel interaction model.**

## S1. Justification of lidocaine-$Na^+$ channel interaction approximations.

In formulating our low-dimensional model of the lidocaine-Na+ channel interaction (Section 2.2 of the main text), we assume that: (1) Charged lidocaine has no effect on $Na^+$ channel kinetics; (2) Neutral lidocaine only binds to and unbinds from inactivated $Na^+$ channels; and (3) Binding of neutral lidocaine locks $Na^+$ channels in the inactivated state until the drug unbinds. Here, we augment the arguments laid out in Section 2.2 of the main text, providing further justification for our modeling assumptions.

### S1.1. Direct simulations of the modified ten Tusscher et al. model demonstrate that lidocaine's effects are due to the neutral form binding to inactivated channels

The above assumptions assert that lidocaine binding only consists of neutral drug binding to inactivated channels. In the main text, we base our assumptions on the rate constants of the Moreno et al. model [1]. To directly assess the validity of these assumptions, we simulated the ten Tusscher et al. model [2, 3] with the Moreno et al. model at a BCL of $750\ ms$ and recorded the fraction of $Na^+$ channels bound to lidocaine and the state of the blocked channels. Figure S1 displays resulting time courses of transmembrane potential (A and C; $5\ \mu M$ and $20\ \mu M$ of lidocaine, respectively) and fraction of channels in inactivated states bound to neutral lidocaine, in non-inactivated states bound to neutral lidocaine, and bound to charged lidocaine (orange, yellow, and purple lines, respectively in panels B and D). At all times, the vast majority of drug bound channels are bound to the neutral form of lidocaine and are in the inactivated state (orange lines). In fact, throughout the $750\ ms$ BCL displayed in Figure S1, the fraction of channels bound to neutral drug in a non-inactivated state or bound to charged drug in any state never exceed $3.8 \times 10^{-3}$ and $4.8 \times 10^{-3}$, respectively, when $[D] = 20\ \mu M$. Moreover, the proportion of drug bound channels that are in an inactivated state and bound to neutral drug never falls below $95\%$ of all drug bound channels for either $[D] = 5\ \mu M$ or $20\ \mu M$.

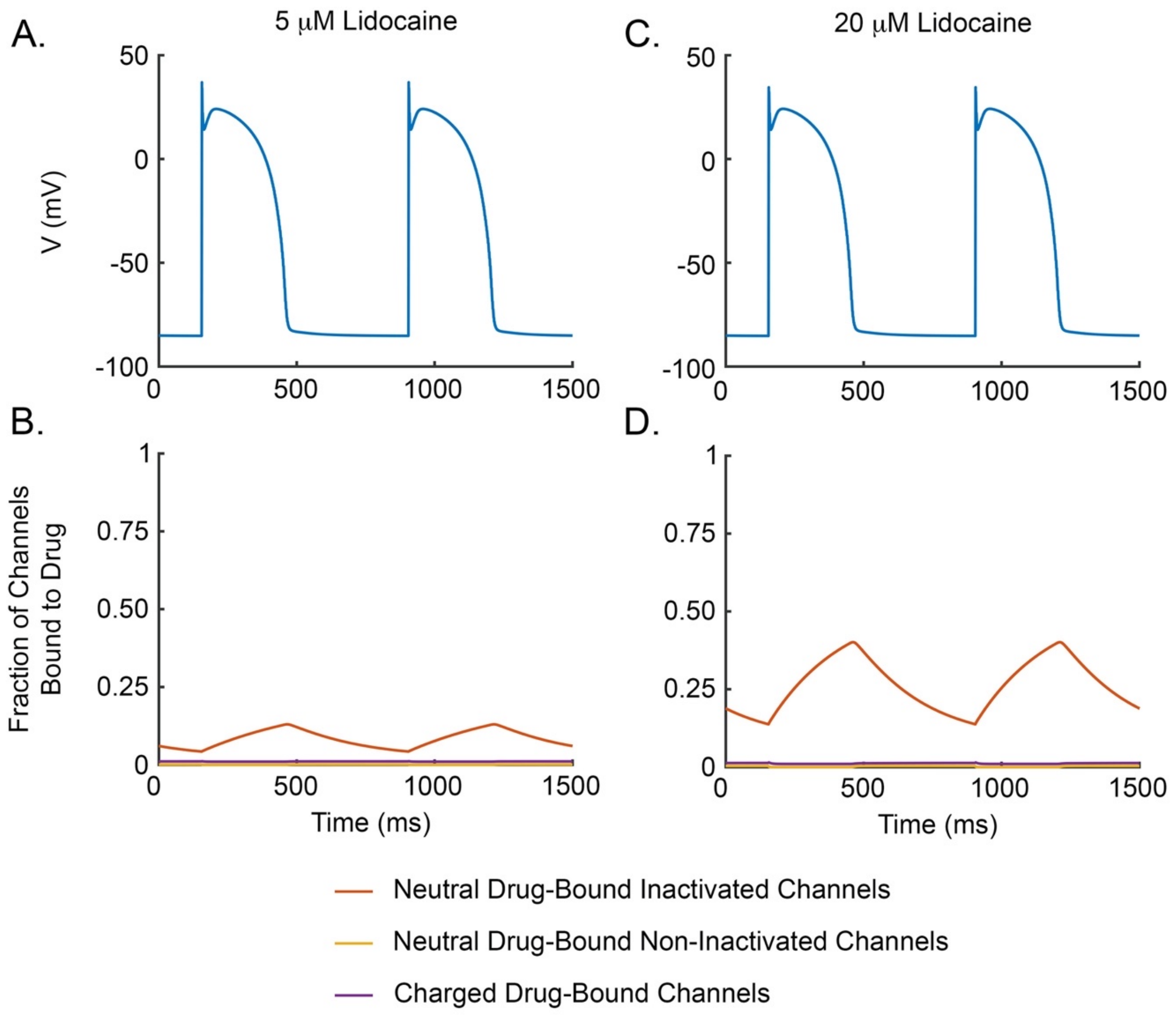

***Figure S1: Fraction of channels bound to drug in Moreno et al. model.*** *Transmembrane potential time course during pacing with* $BCL = 750\ ms$ *for the ten Tusscher et al. model with the Moreno et al. Na⁺ current model in the presence of* $5$ *and* $20\ \mu M$ *of lidocaine (A and C, respectively). The corresponding time courses for fractions of channels bound to neutral drug and in an inactivated state (orange lines), bound to neutral drug and in a non-inactivated state (yellow lines), or bound to charged drug (purple lines) are plotted in (B) and (D). Values for charged drug-bound channels were artificially increased by* $0.01$ *to make lines visible.*

## S1.2. Relative stability analysis indicates that neutral lidocaine stabilizes the inactivated state of the $Na^+$ channel

Further support for our approximation that neutral lidocaine is only ever bound to inactivated channels comes from examining the relative magnitudes of the transition rate constants between inactivated and non-inactivated states when neutral lidocaine is bound. We find that in the Moreno et al. model, neutral drug-bound non-inactivated states are substantially less stable than neutral drug-bound inactivated states, meaning that following neutral drug binding to an inactivated channel, it is highly unlikely that the channel will transition to a non-inactivated state prior to drug unbinding.

To examine inactivation transition rates and stability of neutral drug bound channels (and compare them to that of non-drug bound channels), we define the "relative stability" of inactivated states to be the ratio of the steady state occupancy of non-inactivated states to that of inactivated states. For example, for the transition from the fast inactivation process to the closed states, channels recover from inactivation and inactivate with the rate constants $\alpha 3$ and $\beta 3$, respectively, when no drug is bound and $\alpha_33$ and $\beta_33$ when neutral drug is bound (see Fig 1B in the main text). Therefore, because inactivation and recovery from inactivation are in equilibrium at steady state, $\beta 3 * C = \alpha 3 * I$ and $\beta_33 * DC = \alpha_33 * DI$ where $C$ and $I$ represent the fraction of channels in the $C1$, $C2$, or $C3$ and $IF$, $IC2$, or $IC3$ states, respectively (similar for the drug bound states). Hence,

$$\frac{C}{I} = \frac{\alpha 3}{\beta 3} \quad \text{and} \quad \frac{DC}{DI} = \frac{\alpha_33}{\beta_33}.$$

For the fast and slow inactivation processes from the open state, similar calculations yield

$$\frac{O}{IF}=\frac{\beta 2}{\alpha 2}, \qquad \frac{DO}{DIF}=\frac{\beta_22}{\alpha_22}, \qquad \frac{O}{IS}=\frac{\beta x}{\alpha x}, \text{ and } \frac{DO}{DIS}=\frac{\beta x2}{\alpha x2}.$$

Figure S2A displays that the ratio of closed to fast inactivated states is 265 times smaller when neutral drug is bound ($DC/DI$, orange line) than in the drug-free case ($C/I$, blue line). Similarly, Figure S2B and C display that the ratios of open to fast inactivated states and open to slow inactivated states are 59 times smaller when drug is bound ($DO/DIF$ and D$O/DIS$, orange lines) than in the drug-free state ($O/IF$ and $O/IS$, blue lines). In summary, following neutral lidocaine binding to an inactivated $Na^+$ channel, it is highly unlikely that the channel will transition to a non-inactivated state prior to drug unbinding.

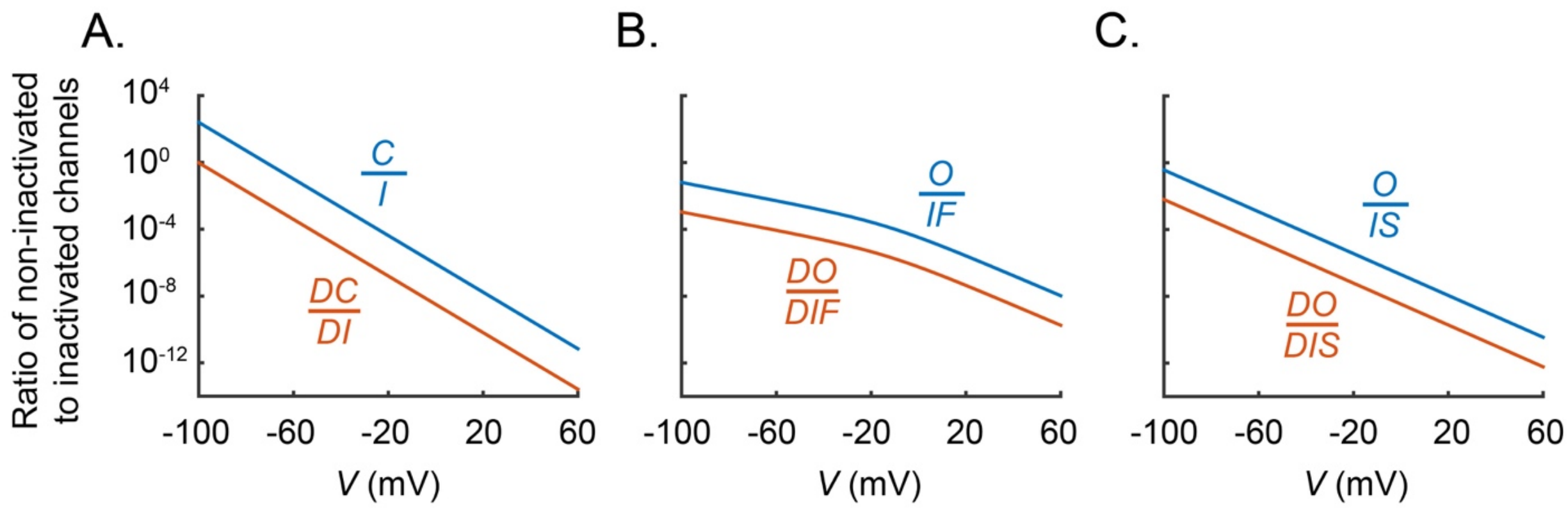

***Figure S2: Inactivation stability.*** *Steady state ratio of fractions of channels in non-inactivated states to inactivated states in the Moreno et al. 2011 [1] lidocaine model as a function of transmembrane potential, $V$. Lines represent the ratios of closed to fast inactivated (A), open to fast inactivated (B), and open to slow inactivated (C) states, for channels with (orange lines) and without (blue lines) neutral drug bound.*

## S2. Experimental voltage-clamp protocols

The following two subsections outline the experimental Voltage-clamp protocols used to generate the data presented in Figs 2 (drug-free data) and 3 (lidocaine data) in the main text.

### S2.1. Drug-free experimental protocols for data in Fig 2 of main text

Here, we briefly describe the experimental protocols used to generate the data presented in Fig 2 of the main text, to which our low-dimensional model was fit.

Steady state availability (Fig 2A): The steady state availability data is from Fig 5B of Liu et al. [1] and was collected from HEK293 cells expressing $Na^+$ channels at 22°C. Conditioning pulses at various potentials ($V_{cond}$) were followed by a test pulse to $-10$ mV, and peak conductance was recorded. Peak conductances corresponding to each $V_{cond}$ were normalized to the peak conductance for to the most negative conditioning pulse ($-130\ mV$).

Steady state activation (Fig 2B): The steady state activation data is from Fig 4A of Rivolta et al. [2] and was collected from HEK293 cells expressing $Na^+$ channels at 22°C. Conditioning pulses at $-100\ mV$ were followed by test pulses to varying potentials, $V_{test}$, and peak conductance was recorded. Peak conductances corresponding to each $V_{test}$ were normalized by the maximal peak conductance observed.

Time constant of inactivation (Fig 2C): (i) For hyperpolarized potentials, time to half-recovery from inactivation is from Supplementary Information of Moreno et al. 2016 [3] that was collected from HEK293 cells expressing $Na^+$ channels at 22°C. Inactivation was induced with a $-10\ mV$ conditioning pulse ($100\ ms$) before holding at the recovery voltage ($V_{recov} = -120\ mV$, $-100\ mV$, or $-80\ mV$). Test pulses to $-10\ mV$ for $25ms$ were applied after various recovery intervals $(\Delta t)$ and peak current recorded. We used this data to approximate the recovery interval $(t_{1/2})$ that would result in a peak conductance that is half the maximal peak conductance (i.e., following an infinite recovery interval). (ii) For depolarized potentials, time to half inactivation is from Fig 3B of Rivolta et al. [2], and was collected from HEK293 cells expressing $Na^+$ channels at $22$°C. Conditioning pulses at $-100\ mV$ were followed by test pulses of $40\ ms$ to various potentials ($V_{test} = -35\ mV$ to $20\ mV$ in $5\ mV$ increments). Time required for the conductance to decay to half its peak value was recorded.

Time constant of activation, $\tau_m$ (Fig 2D): The time constant of activation data is computed from the experimental data in Table 1 of Mitsuiye and Noma [4], which was collected from guinea-pig heart cells at $19$°C [4]. N-bromoacetamide was used to remove $Na^+$ channel inactivation. Conditioning pulses at $-100\ mV$ were followed by test pulses to various potentials and current traces were recorded. Current traces of $Na^+$ current activation were fit with a single exponential. We use a least squares fit to find the $\tau_m$ that gives the best agreement between our $m^3$ activation model and the Mitsuiye and Noma single exponential. The data points in Fig 2D indicate $\tau_m$s from the least squares fit.

It should be noted that Mitsuiye and Noma performed their experiments at $19$°C, whereas all other data in Fig 2 of the main text were collected at $22$°C. Therefore, activation time constant data from [4] was adjusted to $22$°C using a $Q_{10}$ factor of $3$.

### S2.1.1. Differences in data for fitting of drug-free Moreno et al. model and low-dimensional model

There are some slight differences in the data used to fit the Moreno et al. model and our Hodgkin-Huxley model. First, we fit to the activation time constant data in Fig 2D, whereas Moreno et al. used mean channel open time at $-30\ mV$ [5] to constrain activation rates. Second, instead of time to half inactivation for $V = -120, -100$, and $-80\ mV$, which we use for the low-dimensional model, the Moreno et al. model was fit to a time course for recovery from inactivation at $-100\ mV$ to constrain the time constant of inactivation at hyperpolarized potentials. Third, the Moreno et al. model was only fit to time to half inactivation data for $V$ between $-20$ and $20\ mV$, while we use all data for $V$ between $-35$ and $20\ mV$ shown in Fig 2C. Finally, the Moreno et al. model was also fit to recovery from use dependent block data, which provides information about the slow inactivation processes. However, we purposefully chose the simplest Hodgkin-Huxley type $Na^+$ current model, which does not include a slow inactivation gate (often referred to as a "j gate"). Therefore, we excluded the recovery from use dependent block data from our optimization.

## S2.2. Lidocaine experimental protocols for data in Fig 3 of main text

All experimental data characterizing the effects of lidocaine were collected using voltage-clamp experiments performed at $22°C$ on HEK293 cells expressing only $Na^+$ channels [1, 5, 6].

Steady state availability (Fig 3A): The steady state availability data is from Fig 5B of Liu et al. [1]. After the addition of $100\ \mu M$ of lidocaine, cells were held at $-100\ mV$ for $10\ s$ prior to $5\ s$ conditioning pulses at various potentials. Peak conductances were measured during a $25\ ms$ test pulse at $-10\ mV$. Peak conductance data was normalized by the peak conductance resulting from the most negative conditioning pulse ($-130\ mV$).

Frequency dependence of block (Fig 3B): The frequency dependent block data is from Fig 3C of Abriel et al. [6]. After the addition of $300\ \mu M$ of lidocaine, cells were paced $100$ times from $-100\ mV$ to $-10\ mV$ for $25\ ms$ at various frequencies. The differences between the peak conductance during the first and last stimulus were calculated and then normalized by the peak conductance during the first stimulus to obtain fractional block.

Tonic block (Fig 3C): The tonic block data is from Fig 2C of Abriel et al. [6]. Cells were held at $-100\ mV$ with various concentrations of lidocaine. Peak conductance was measured during a $25\ ms$ test pulse to $-10\ mV$. Peak conductances were then normalized to the peak conductance in the absence of drug.

Dose dependence of use-dependent block (Fig 3D): The dose dependence of use-dependent block data is from Fig 3A of Abriel et al. [6]. Cells were paced $600$ times from $-100\ mV$ to $-10\ mV$ for $25\ ms$, at a frequency of $5\ Hz$. Peak conductance during the last stimulus was normalized by the peak conductance elicited from a cell held at $-100\ mV$ and then stimulated to $-10\ mV$ in the absence of drug. Note: in the Fig in Abriel et al., peak conductance during the last stimulus is normalized by tonic block.

Recovery from use-dependent block (Fig 3E): The recovery from use-dependent block data is from Figs 6A in Liu et al. [5] and 1C in Moreno et al. [7] for drug-free and in the presence of $300\ \mu M$ lidocaine, respectively. Cells were paced $100$ times from $-100\ mV$ to $-10\ mV$ for $25\ ms$, at a frequency of $25\ Hz$. Cells were then held at $-100\ mV$ for variable intervals before a test pulse to $-10\ mV$ was applied and peak conductance measured. Peak conductance was normalized by peak conductance during slow pacing at $0.033\ Hz$.

## S3. Low-dimensional model under predicts tonic block at $-100\ mV$ (but not at the more physiological resting potential of $-85\ mV$)

Our low-dimensional model under predicts the level of tonic block at $-100\ mV$ (Fig 3C). On the other hand, the Moreno et al. model accurately reproduces the experimental tonic block data. Here, we use the Moreno et al. model to explain why our model under predicts tonic block at $-100\ mV$. We also demonstrate that our model agrees with the Moreno et al. model at a more physiological resting potential ($-85\ mV$), and therefore presumably our model replicates the physiological level of tonic block.

The disparity between the experimental data and our low-dimensional model's predictions of tonic block (Fig 3C) can be understood by considering the $K_d$'s (ratios of drug unbinding rate to drug binding rate) for lidocaine-$Na^+$ channel interactions in various conformational states. As stated in Section 2.2, neutral lidocaine binds to inactivated $Na^+$ channels with a $K_d$ of 6.8 $\mu M$ [1, 2]. However, at $V = -100\ mV$, very few $Na^+$ channels are inactivated. Therefore, because our low-dimensional model includes only the neutral form of lidocaine binding to inactivated channels, the overall binding and unbinding rates at $V = -100\ mV$ are $\left(1 - h_\infty(-100)\right)k_{on}$ and $k_{off}$, respectively, and the $K_d$ is

$$K_d(-100) = \frac{k_{off}}{\left(1 - h_\infty(-100)\right)k_{on}} = 1100\ \mu M.$$

This $K_d$ at $-100\ mV$ is the same order of magnitude as the $K_d$'s for neutral lidocaine binding to non-inactivated channels (1800 and 400 $\mu M$ for closed and open channels, respectively) and charged lidocaine binding to non-inactivated channels at $-100\ mV$ ($5000\ \mu M$) at 22 °C [1-4]. Hence, at $-100\ mV$ lidocaine binding is not dominated by the neutral form binding to inactivated channels, as we assume in our low-dimensional model.

The Moreno et al. model replicates tonic block data better than our model (Figure S3A) because, unlike our model, it includes the effects of charged lidocaine and the ability of lidocaine to bind to non-inactivated channels. Specifically, predicted tonic block decreases (i.e., normalized peak conductance increases) in the Moreno et al. model when the effects of charged lidocaine are removed from the model (dashed yellow line), and decreases further when neutral drug-bound, non-inactivated states are also removed from the model (dashed purple line).

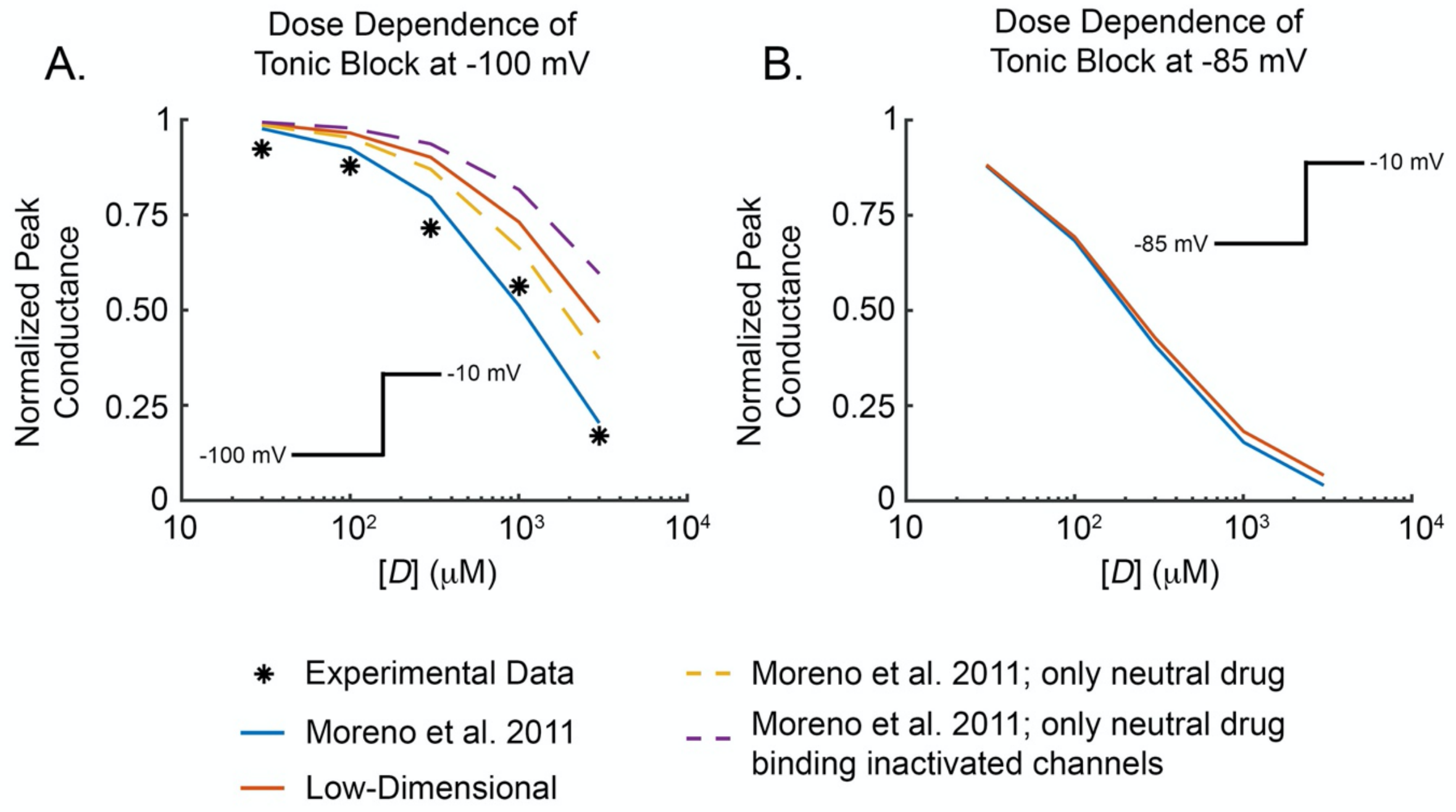

***Figure S3: Tonic block with holding potentials of $-100\ mV$ (A) and $-85\ mV$ (B).*** *Experimental data (asterisks), Moreno et al. 2011 model (blue lines), low-dimensional model (orange lines), Moreno et al. 2011 model with only neutral drug effects (dashed yellow line), and Moreno et al. 2011 model with only neutral drug binding to inactivated channels (dashed purple line).*

However, predictions for tonic block at a more physiological resting potential ($-85\ mV$) from our low-dimensional model and the Moreno et al. model nearly overlap (Figure S3B). In fact, the largest observed difference in normalized peak conductance is $0.031$ at a lidocaine concentration of $1000\ \mu M$. The improved agreement in tonic block at less hyperpolarized potentials is due to the greater fraction of inactivated $Na^+$ channels, which causes lidocaine binding to be dominated by the neutral form binding to inactivated channels.

## S4. Peak upstroke and conduction velocity in modified ten Tusscher et al. models

Fig 4 of the main text shows the dependencies of normalized peak upstroke velocity and conduction velocity on BCL in the ten Tusscher et al. human ventricular myocyte model [1, 2] with the Moreno et al. [3] and our low-dimensional models of the $Na^+$ current. To construct the modified ten Tusscher et al. models, we replace the ten Tusscher fast $Na^+$ conductance with either the Moreno et al. model or our low-dimensional model. As the ten Tusscher et al. model is parameterized for $37°C$, we use the $37°C$ parameterizations of the Moreno et al. and low-dimensional models. The maximal $Na^+$ conductance per capacitance in the Moreno et al. model is $15\ nS/pF$ [3], and the maximal conductance per capacitance of the low-dimensional model is set to $20\ nS/pF$, so that conduction velocity at a BCL of $1000\ ms$ with no drug is the same in both models.

For each BCL, peak upstroke velocity is recorded after the model cell is paced $500$ times with stimuli of amplitude $-80\ pA/pF$ and duration $1\ ms$.

To measure conduction velocity, waves are generated by depolarizing the distal $0.1\ cm$ of a $1.1\ cm$ long cable to $0\ mV$. $500$ waves are generated at a BCL of $1000\ ms$ and the BCL is subsequently decreased to $300\ ms$ in increments of $50\ ms$ with $10$ stimuli applied at each BCL. Conduction velocity is recorded for the last two stimuli of each pacing rate at the point $0.6\mathrm{cm}$ from the distal end.

Normalized upstroke and conduction velocities for the modified ten Tusscher et al. models with $5\ \mu M$ and $20\ \mu M$ of lidocaine are presented in Fig 4; here, the corresponding non-normalized upstroke and conduction velocities are displayed in Figure S4 along with the results for the drug-free cases. The maximal $Na^+$ conductance in the low-dimensional model (conductance per capacitance of $20\ nS/pF$) is relatively high compared to the Moreno et al. model at (conductance per capacitance of $15\ nS/pF$), causing peak upstroke velocity to be greater in the low-dimensional model than in the Moreno et al. model. Note that the rate-dependencies of peak upstroke velocity and conduction velocity are slightly different in the two drug-free models, with upstroke and conduction velocities decreasing more at short BCL in the Moreno et al. model. The disparities in the rate-dependencies of the drug-free models are due to the differences in formulation of the drug-free $Na^+$ current models, which are also responsible for some of the discrepancy in rate-dependence of the models in the presence of lidocaine. Therefore, in the main text, we present upstroke and conduction velocities normalized by peak upstroke and conduction velocities of the corresponding drug-free model, so as to focus on the rate-dependent effects of lidocaine and not the combined rate-dependence of lidocaine and the underlying $Na^+$ current.

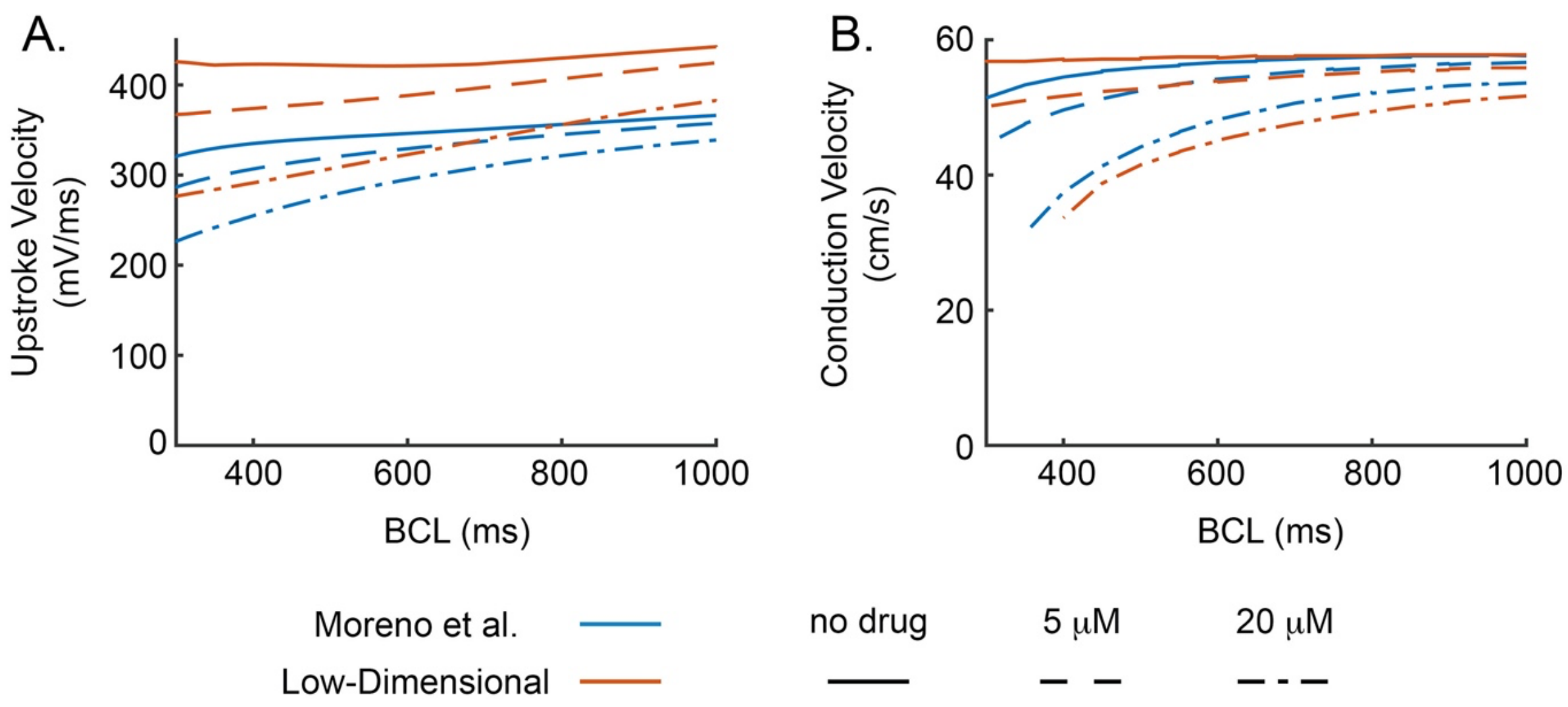

***Figure S4: Rate-Dependent Effects of Lidocaine.*** *Peak upstroke velocity (A) and conduction velocity (B) plotted against BCL for the ten Tusscher et al. human ventricular myocyte model [1, 2] with the Moreno et al. model (blue lines) or low-dimensional model (orange lines) of the* $Na^+$ *current. Peak upstroke and conduction velocities in the absence of drug (solid lines), and in the presence of* $5\ \mu M$ *(dashed lines) and* $20\ \mu M$ *(dot-dashed lines) concentrations of lidocaine are plotted.*

## S5. Dependence of lidocaine binding on physiological properties

In Section 3.2.2 of the main text, we derive an expression for the fraction of channels bound to neutral lidocaine during the upstroke, $b^*$, that captures the parametric dependence on BCL, action potential duration restitution properties, plateau potential, diastolic potential, channel inactivation kinetics, and drug concentration and binding rates. We then use this expression for $b^*$ and the partial derivatives of $b^*$ to examine how lidocaine binding depends on the physiological properties of restitution curve and transmembrane potential during the AP and DI. Here, we derive the partial derivatives of $b^*$.

The expression for $b^*$ is

$$b^* = \left[\frac{1-D}{1-AD}\right] b_\infty(V_{DI}) + \left[\frac{(1-A)D}{1-AD}\right] b_\infty(V_{AP}),$$

where

$$A = e^{-\frac{APD}{\tau_b(V_{AP})}}, \qquad D = e^{-\frac{DI}{\tau_b(V_{DI})}}$$

and

$$\tau_b(V) = \frac{1}{\big(1-h_\infty(V)\big)[D]k_{on} + k_{off}}, \qquad b_\infty(V) = \frac{\big(1-h_\infty(V)\big)[D]k_{on}}{\big(1-h_\infty(V)\big)[D]k_{on} + k_{off}}.$$

Also, we define $APD = f(BCL)$ to be the restitution curve and note that $DI = BCL - APD = BCL - f(BCL)$.

**S5.1. Partial derivative of $b^*$ with respect to BCL**

$$\frac{\partial b^*}{\partial BCL} = \left[-\frac{\partial D}{\partial BCL}(1-AD)^{-1} + (1-D)(1-AD)^{-2}\left(A\frac{\partial D}{\partial BCL} + D\frac{\partial A}{\partial BCL}\right)\right] b_\infty(V_{DI})$$

$$+ \left[\frac{\partial D}{\partial BCL}(1-A)(1-AD)^{-1} - \frac{\partial A}{\partial BCL}D(1-AD)^{-1}\right.$$

$$\left. + (1-A)D(1-AD)^{-2}\left(A\frac{\partial D}{\partial BCL} + D\frac{\partial A}{\partial BCL}\right)\right] b_\infty(V_{AP})$$

$$= \left[-\frac{\partial D}{\partial BCL}(1-AD) + (1-D)\left(A\frac{\partial D}{\partial BCL} + D\frac{\partial A}{\partial BCL}\right)\right]\frac{b_\infty(V_{DI})}{(1-AD)^2}$$

$$+ \left[\frac{\partial D}{\partial BCL}(1-A)(1-AD) - \frac{\partial A}{\partial BCL}D(1-AD)\right.$$

$$\left. + (1-A)D\left(A\frac{\partial D}{\partial BCL} + D\frac{\partial A}{\partial BCL}\right)\right]\frac{b_\infty(V_{AP})}{(1-AD)^2}$$

$$= \left[(A-1)\frac{\partial D}{\partial BCL} + D(1-D)\frac{\partial A}{\partial BCL}\right]\frac{b_\infty(V_{DI})}{(1-AD)^2}$$

$$+ \left[(1-A)\frac{\partial D}{\partial BCL} + D(D-1)\frac{\partial A}{\partial BCL}\right]\frac{b_\infty(V_{AP})}{(1-AD)^2}$$

$$= \frac{1}{(1-AD)^2}\left[b_\infty(V_{AP}) - b_\infty(V_{DI})\right]\left[(1-A)\frac{\partial D}{\partial BCL} - D(1-D)\frac{\partial A}{\partial BCL}\right].$$

We note that

$$\frac{\partial D}{\partial BCL} = \frac{\partial}{\partial BCL}\left(e^{-\frac{DI}{\tau_b(V_{DI})}}\right)$$

$$= -\frac{1}{\tau_b(V_{DI})}\frac{dDI}{dBCL}e^{-\frac{DI}{\tau_b(V_{DI})}}$$

$$= -\frac{1}{\tau_b(V_{DI})}\left(1 - \frac{df(BCL)}{dBCL}\right)e^{-\frac{DI}{\tau_b(V_{DI})}}$$

$$= -\frac{1}{\tau_b(V_{DI})}\left(1 - \frac{df(BCL)}{dBCL}\right)D$$

and

$$\frac{\partial A}{\partial BCL} = \frac{\partial}{\partial BCL}\left(e^{-\frac{APD}{\tau_b(V_{AP})}}\right)$$

$$= -\frac{1}{\tau_b(V_{AP})}\frac{dAPD}{dBCL}e^{-\frac{APD}{\tau_b(V_{AP})}}$$

$$= -\frac{1}{\tau_b(V_{AP})}\frac{df(BCL)}{dBCL}e^{-\frac{APD}{\tau_b(V_{AP})}}$$

$$= -\frac{1}{\tau_b(V_{AP})}\frac{df(BCL)}{dBCL}A.$$

Inserting the expressions for $\frac{\partial D}{\partial BCL}$ and $\frac{\partial A}{\partial BCL}$ into the expression for $\frac{\partial b^*}{\partial BCL}$ produces

$$\frac{\partial b^*}{\partial BCL} = \frac{b_\infty(V_{AP}) - b_\infty(V_{DI})}{(1 - AD)^2}\left[\frac{(1 - D)AD}{\tau_b(V_{AP})}\frac{df(BCL)}{dBCL} - \frac{(1 - A)D}{\tau_b(V_{DI})}\left(1 - \frac{df(BCL)}{dBCL}\right)\right],$$

which is Equation (6) in the main text.

**S5.2. Partial derivative of $b^*$ with respect to $V_{AP}$**

$$\frac{\partial b^*}{\partial V_{AP}} = (1-D)(1-AD)^{-2}\frac{\partial A}{\partial V_{AP}}Db_\infty(V_{DI})$$

$$+\left[-\frac{\partial A}{\partial V_{AP}}D(1-AD)^{-1} + (1-A)D(1-AD)^{-2}\frac{\partial A}{\partial V_{AP}}D\right]b_\infty(V_{AP})$$

$$+(1-A)D(1-AD)^{-1}\frac{\partial b_\infty(V_{AP})}{\partial V_{AP}}$$

$$=\frac{1}{(1-AD)^2}[(1-D)Db_\infty(V_{DI}) + [-(D-AD^2)+(D^2-AD^2)]b_\infty(V_{AP})]\frac{\partial A}{\partial V_{AP}}$$

$$+(1-A)D(1-AD)^{-1}\frac{\partial b_\infty(V_{AP})}{\partial V_{AP}}$$

$$=\frac{1}{(1-AD)^2}[(1-D)Db_\infty(V_{DI}) + (D-1)Db_\infty(V_{AP})]\frac{\partial A}{\partial V_{AP}}$$

$$+(1-A)D(1-AD)^{-1}\frac{\partial b_\infty(V_{AP})}{\partial V_{AP}}$$

$$=\frac{(1-D)D}{(1-AD)^2}[b_\infty(V_{DI}) - b_\infty(V_{AP})]\frac{\partial A}{\partial V_{AP}} + \frac{(1-A)D}{1-AD}\frac{\partial b_\infty(V_{AP})}{\partial V_{AP}}.$$

We note that

$$A = e^{-\frac{APD}{\tau_b(V_{AP})}} = e^{-APD\left[\left(1-h_\infty(V_{AP})\right)[D]k_{on}+k_{off}\right]}$$

and

$$b_\infty(V_{AP}) = \frac{\big(1 - h_\infty(V_{AP})\big)[D]k_{on}}{\big(1 - h_\infty(V_{AP})\big)[D]k_{on} + k_{off}},$$

therefore,

$$\frac{\partial A}{\partial V_{AP}} = \frac{\partial}{\partial V_{AP}}\left(e^{-APD\left[\left(1-h_\infty(V_{AP})\right)[D]k_{on}+k_{off}\right]}\right)$$

$$= APD\frac{dh_\infty(V_{AP})}{dV_{AP}}[D]k_{on}e^{-APD\left[\left(1-h_\infty(V_{AP})\right)[D]k_{on}+k_{off}\right]}$$

$$= APD[D]k_{on}\frac{dh_\infty(V_{AP})}{dV_{AP}}A$$

and

$$\frac{\partial b_\infty(V_{AP})}{\partial V_{AP}} = -\frac{dh_\infty(V_{AP})}{dV_{AP}}\frac{[D]k_{on}}{\big(1 - h_\infty(V_{AP})\big)[D]k_{on} + k_{off}}$$

$$+ \big(1 - h_\infty(V_{AP})\big)[D]k_{on}\left(\big(1 - h_\infty(V_{AP})\big)[D]k_{on} + k_{off}\right)^{-2}\frac{dh_\infty(V_{AP})}{dV_{AP}}[D]k_{on}$$

$$= -\tau_b(V_{AP})[D]k_{on}\frac{dh_\infty(V_{AP})}{dV_{AP}} + b_\infty(V_{AP})\tau_b(V_{AP})[D]k_{on}\frac{dh_\infty(V_{AP})}{dV_{AP}}$$

$$= \tau_b(V_{AP})[D]k_{on}(b_\infty(V_{AP}) - 1)\frac{dh_\infty(V_{AP})}{dV_{AP}}.$$

Inserting the expressions for $\frac{\partial A}{\partial V_{AP}}$ and $\frac{\partial b_\infty(V_{AP})}{\partial V_{AP}}$ into the expression for $\frac{\partial b^*}{\partial V_{AP}}$ produces

$$\frac{\partial b^*}{\partial V_{AP}} = \frac{(1-D)AD}{(1-AD)^2}[b_\infty(V_{DI}) - b_\infty(V_{AP})]APD[D]k_{on}\frac{dh_\infty(V_{AP})}{dV_{AP}}$$

$$+ \frac{(1-A)D(1-AD)}{(1-AD)^2}\tau_b(V_{AP})[D]k_{on}[b_\infty(V_{AP}) - 1]\frac{dh_\infty(V_{AP})}{dV_{AP}}$$

$$= \frac{D[D]k_{on}}{(1-AD)^2}\big[(1-D)A[b_\infty(V_{AP}) - b_\infty(V_{DI})]APD$$

$$+ (1-A)(1-AD)\tau_b(V_{AP})[1 - b_\infty(V_{AP})]\big]\frac{d\big(1-h_\infty(V_{AP})\big)}{dV_{AP}}$$

$$= \xi(V_{AP})\ \frac{d\big(1-h_\infty(V_{AP})\big)}{dV_{AP}},$$

where

$$\xi(V_{AP}) = \frac{D\tau_b(V_{AP})[D]k_{on}}{(1-AD)^2}\left[A(1-D)\frac{APD}{\tau_b(V_{AP})}[b_\infty(V_{AP}) - b_\infty(V_{DI})]\right.$$

$$\left. + (1-A)(1-AD)[1-b_\infty(V_{AP})]\right].$$

Note that $A, D < 1$, and $(1-A), (1-D), (1-AD) > 0$, and $1 > b_\infty(V_{AP}) > b_\infty(V_{DI}) > 0$, thus $\xi(V_{AP}) > 0$. Note further that $(1-A), (1-D) \le (1-AD)$ and $(1-A) \ge \frac{APD}{\tau_b(V_{AP})}A$ because $1 - e^{-x} \ge xe^{-x}$ for all $x \ge 0$, thus

$$\xi(V_{AP}) < \frac{D\tau_b(V_{AP})[D]k_{on}}{(1-AD)^2}\big[(1-A)(1-AD)[1-b_\infty(V_{DI})]$$

$$+ (1-A)(1-AD)[1-b_\infty(V_{AP})]\big]$$

$$< \frac{D}{(1-AD)}\left[(1-AD)[1-b_\infty(V_{DI})] + (1-AD)[1-b_\infty(V_{DI})]\right]\tau_b(V_{AP})[D]k_{on}$$

$$= 2D[1-b_\infty(V_{DI})]\ \frac{[D]k_{on}}{\big(1-h_\infty(V_{AP})\big)[D]k_{on} + k_{off}}$$

$$< 2\frac{[D]k_{on}}{k_{off}} = 2\frac{[D]}{K_d}.$$

For the parameters considered here, $K_d = 6.8\ \mu M$, and at clinical concentrations, the concentration of neutral lidocaine is $[D] < 8\ \mu M$. Hence, $0 < \xi(V_{AP}) < 2.4$.

Therefore, because steady state inactivation is saturated above $-50\ mV$ (i.e., $\frac{d(1-h_\infty(V_{AP}))}{dV_{AP}} \leq 4.5 \times 10^{-3}\ mV^{-1}$ for $V_{AP} > -50\ mV$ at 37°C), $b^*$ is insensitive to changes in $V_{AP}$ (i.e., $\frac{\partial b^*}{\partial V_{AP}} \approx 0$).

**S5.3. Partial derivative of $b^*$ with respect to $V_{DI}$**

$$\frac{\partial b^*}{\partial V_{DI}} = \left[-\frac{\partial D}{\partial V_{DI}}(1-AD)^{-1} + (1-D)(1-AD)^{-2}A\frac{\partial D}{\partial V_{DI}}\right]b_\infty(V_{DI}) + \frac{1-D}{1-AD}\frac{\partial b_\infty(V_{DI})}{\partial V_{DI}}$$

$$+\left[(1-A)(1-AD)^{-1}\frac{\partial D}{\partial V_{DI}} + (1-A)D(1-AD)^{-2}A\frac{\partial D}{\partial V_{DI}}\right]b_\infty(V_{AP})$$

$$= \frac{1}{(1-AD)^2}\big[[-(1-AD) + A(1-D)]b_\infty(V_{DI})$$

$$+ [(1-A)(1-AD) + (1-A)AD]b_\infty(V_{AP})\big]\frac{\partial D}{\partial V_{DI}} + \frac{1-D}{1-AD}\frac{\partial b_\infty(V_{DI})}{\partial V_{DI}}$$

$$= \frac{1}{(1-AD)^2}[(A-1)b_\infty(V_{DI}) + (1-A)b_\infty(V_{AP})]\frac{\partial D}{\partial V_{DI}} + \frac{1-D}{1-AD}\frac{\partial b_\infty(V_{DI})}{\partial V_{DI}}$$

$$= \frac{1-A}{(1-AD)^2}\left(b_\infty(V_{AP}) - b_\infty(V_{DI})\right)\frac{\partial D}{\partial V_{DI}} + \frac{1-D}{1-AD}\frac{\partial b_\infty(V_{DI})}{\partial V_{DI}}.$$

We note that similar to $A$ and $b_\infty(V_{AP})$,

$$D = e^{-\frac{DI}{\tau_b(V_{DI})}} = e^{-DI\left[\left(1-h_\infty(V_{DI})\right)[D]k_{on}+k_{off}\right]}$$

and

$$b_\infty(V_{DI}) = \frac{\left(1-h_\infty(V_{DI})\right)[D]k_{on}}{\left(1-h_\infty(V_{DI})\right)[D]k_{on}+k_{off}},$$

so by the calculations done in Section S5.2,

$$\frac{\partial D}{\partial V_{DI}} = DI[D]k_{on}\frac{dh_\infty(V_{DI})}{dV_{DI}}D$$

and

$$\frac{\partial b_\infty(V_{DI})}{\partial V_{DI}} = \tau_b(V_{DI})[D]k_{on}(b_\infty(V_{DI})-1)\frac{dh_\infty(V_{DI})}{dV_{DI}}.$$

Inserting the expressions for $\frac{\partial D}{\partial V_{DI}}$ and $\frac{\partial b_\infty(V_{DI})}{\partial V_{DI}}$ into the expression for $\frac{\partial b^*}{\partial V_{DI}}$ produces

$$\frac{\partial b^*}{\partial V_{DI}} = \frac{(1-A)D}{(1-AD)^2}\big(b_\infty(V_{AP}) - b_\infty(V_{DI})\big)DI[D]k_{on}\frac{dh_\infty(V_{DI})}{dV_{DI}}$$

$$+\frac{1-D}{1-AD}\tau_b(V_{DI})[D]k_{on}(b_\infty(V_{DI})-1)\frac{dh_\infty(V_{DI})}{dV_{DI}}$$

$$= \frac{[D]k_{on}}{(1-AD)^2}\big[(1-A)D\big(b_\infty(V_{AP}) - b_\infty(V_{DI})\big)DI$$

$$+(1-AD)(1-D)\tau_b(V_{DI})(b_\infty(V_{DI})-1)\big]\frac{dh_\infty(V_{DI})}{dV_{DI}}$$

$$= \frac{\tau_b(V_{DI})[D]k_{on}}{(1-AD)^2}\Big[(1-A)D\frac{-DI}{\tau_b(V_{DI})}\big(b_\infty(V_{AP}) - b_\infty(V_{DI})\big)$$

$$+(1-AD)(1-D)\big(1-b_\infty(V_{DI})\big)\Big]\frac{d\big(1-h_\infty(V_{DI})\big)}{dV_{DI}}$$

$$= \gamma(V_{DI})\frac{d\big(1-h_\infty(V_{DI})\big)}{dV_{DI}},$$

where

$$\gamma(V_{DI}) = \frac{\tau_b(V_{DI})[D]k_{on}}{(1-AD)^2}\Big[(1-A)D\frac{-DI}{\tau_b(V_{DI})}\big(b_\infty(V_{AP}) - b_\infty(V_{DI})\big)$$

$$+(1-AD)(1-D)\big(1-b_\infty(V_{DI})\big)\Big]$$

Note that $0 < 1-A < 1-AD$, and $0 < b_\infty(V_{AP}) - b_\infty(V_{DI}) < 1 - b_\infty(V_{DI})$, and $0 < \frac{DI}{\tau_b(V_{DI})}D \leq 1-D$ because $xe^{-x} \leq 1-e^{-x}$ for all $x \geq 0$. Thus, $(1-A)D\frac{DI}{\tau_b(V_{DI})}\big(b_\infty(V_{AP}) -$ $b_\infty(V_{DI})\big) < (1 - AD)(1 - D)\big(1 - b_\infty(V_{DI})\big)$, and as a result $\gamma(V_{DI}) > 0$. Additionally, note that $0 < 1 - D < 1 - AD$, thus

$$\gamma(V_{DI}) < \frac{\tau_b(V_{DI})[D]k_{on}}{(1 - AD)^2}\big[(1 - AD)(1 - D)\big(1 - b_\infty(V_{DI})\big)\big]$$

$$< \frac{\tau_b(V_{DI})[D]k_{on}}{(1 - AD)^2}(1 - AD)^2$$

$$= \frac{[D]k_{on}}{\big(1 - h_\infty(V_{DI})\big)[D]k_{on} + k_{off}} < \frac{[D]}{K_d}.$$

Again, for the parameters considered here, $K_d = 6.8\ \mu M$, and at clinical concentrations, the concentration of neutral lidocaine is $[D] < 8\ \mu M$, so $0 < \gamma(V_{DI}) < 1.2$. Hence, similar to the dependence on $V_{AP}$, $b^*$ is insensitive to shifts in $V_{DI}$ at potentials where steady state inactivation is flat (e.g., $\frac{d\left(1-h_\infty(V_{DI})\right)}{dV_{DI}} < 1.3 \times 10^{-2}\ mV^{-1}$, for $V_{DI} < -85\ mV$ at 37°C). However, $b^*$ is highly sensitive to shifts in $V_{DI}$ at potentials where $\frac{d\left(1-h_\infty(V_{DI})\right)}{dV_{DI}} > 0$ (i.e., $V_{DI} > -85\ mV$).

## S6. Moreno et al. model

The equations for the Moreno et al. [1] $Na^+$ current model are provided below. Rate constants with units of $ms^{-1}$ are for 37°C and state transition rates are adjusted for changes in temperature using a $Q_{10}$ factor of 3 (lidocaine binding rates are independent of temperature unless stated explicitly).

### S6.1. Drug-free Moreno et al. $Na^+$ current model:

$$IS = 1 - (O + C1 + C2 + C3 + IC3 + IC2 + IF)$$

$$\frac{dO}{dt} = \beta x \cdot IS + \alpha 13 \cdot C1 + \beta 2 \cdot IF - (\alpha x + \beta 13 + \alpha 2) \cdot O$$

$$\frac{dC1}{dt} = \beta 13 \cdot O + \alpha 3 \cdot IF + \alpha 12 \cdot C2 - (\alpha 13 + \beta 3 + \beta 12) \cdot C1$$

$$\frac{dC2}{dt} = \beta 12 \cdot C1 + \alpha 3 \cdot IC2 + \alpha 11 \cdot C3 - (\alpha 12 + \beta 3 + \beta 11) \cdot C2$$

$$\frac{dC3}{dt} = \beta 11 \cdot C2 + \alpha 3 \cdot IC3 - (\alpha 11 + \beta 3) \cdot C3$$

$$\frac{dIC3}{dt} = \beta 3 \cdot C3 + \beta 11 \cdot IC2 - (\alpha 3 + \alpha 11) \cdot IC3$$

$$\frac{dIC2}{dt} = \alpha 11 \cdot IC3 + \beta 3 \cdot C2 + \beta 12 \cdot IF - (\beta 11 + \alpha 3 + \alpha 12) \cdot IC2$$

$$\frac{dIF}{dt} = \alpha 12 \cdot IC2 + \beta 3 \cdot C1 + \alpha 2 \cdot O - (\beta 12 + \alpha 3 + \beta 2) \cdot IF$$

with rate constants

$$\alpha 11 = \frac{8.5539}{(7.4392 \times 10^{-2})e^{\frac{-V}{17.0}} + (2.0373 \times 10^{-1})e^{\frac{-V}{150}}}$$

$$\alpha 12 = \frac{8.5539}{(7.4392 \times 10^{-2})e^{\frac{-V}{15.0}} + (2.0373 \times 10^{-1})e^{\frac{-V}{150}}}$$

$$\alpha 13 = \frac{8.5539}{(7.4392 \times 10^{-2})e^{\frac{-V}{12.0}} + (2.0373 \times 10^{-1})e^{\frac{-V}{150}}}$$

$$\beta 11 = (7.5215 \times 10^{-2})e^{\frac{-V}{20.3}}$$

$$\beta 12 = 2.7574e^{\frac{-(V-5)}{20.3}}$$

$$\beta 13 = (4.7755 \times 10^{-1}) * e^{\frac{-(V-10)}{20.3}}$$

$$\alpha 3 = (5.1458 \times 10^{-6})e^{\frac{-V}{8.2471}}$$

$$\beta 3 = 6.1205e^{\frac{V}{13.542}}$$

$$\alpha 2 = 13.370e^{\frac{V}{43.749}}$$

$$\beta 2 = \frac{\alpha 13 * \alpha 2 * \alpha 3}{\beta 13 * \beta 3}$$

$$\alpha x = (3.4229 \times 10^{-2})\alpha 2$$

$$\beta x = (1.7898 \times 10^{-2})\alpha 3$$

**S6.2. Moreno et al. model of lidocaine-Na$^+$ channel interaction:**

Non-drug-bound states:

$$IS = 1 - (O + C1 + C2 + C3 + IC3 + IC2 + IF + D^{+}O + D^{+}C1 + D^{+}C2 + D^{+}C3 + D^{+}IC3 + D^{+}IC2 + D^{+}IF + D^{+}IS + DO + DC1 + DC2 + DC3 + DIC3 + DIC2 + DIF + DIS)$$

$$\frac{dO}{dt} = \beta x \cdot IS + \alpha 13 \cdot C1 + \beta 2 \cdot IF + koff \cdot D^{+}O + k_off \cdot DO - (\alpha x + \beta 13 + \alpha 2 + kon + k_on) \cdot O$$

$$\frac{dC1}{dt} = \beta 13 \cdot O + \alpha 3 \cdot IF + \alpha 12 \cdot C2 + kcoff \cdot D^{+}C1 + kc_off \cdot DC1 - (\alpha 13 + \beta 3 + \beta 12 + kcon + kc_on) \cdot C1$$

$$\frac{dC2}{dt} = \beta 12 \cdot C1 + \alpha 3 \cdot IC2 + \alpha 11 \cdot C3 + kcoff \cdot D^{+}C2 + kc_off \cdot DC2 - (\alpha 12 + \beta 3 + \beta 11 + kcon + kc_on) \cdot C2$$

$$\frac{dC3}{dt} = \beta 11 \cdot C2 + \alpha 3 \cdot IC3 + kcoff \cdot D^{+}C3 + kc_off \cdot DC3 - (\alpha 11 + \beta 3 + kcon + kc_on) \cdot C3$$

$$\frac{dIC3}{dt} = \beta 3 \cdot C3 + \beta 11 \cdot IC2 + ki_off \cdot DIC3 - (\alpha 3 + \alpha 11 + ki_on) \cdot IC3$$

$$\frac{dIC2}{dt} = \alpha 11 \cdot IC3 + \beta 3 \cdot C2 + \beta 12 \cdot IF + ki_off \cdot DIC2 - (\beta 11 + \alpha 3 + \alpha 12 + ki_on) \cdot IC2$$

$$\frac{dIF}{dt} = \alpha 12 \cdot IC2 + \beta 3 \cdot C1 + \alpha 2 \cdot O + ki_off \cdot DIF - (\beta 12 + \alpha 3 + \beta 2 + ki_on) \cdot IF$$

Charged drug-bound states:

$$\frac{dD^{+}O}{dt} = \beta x1 \cdot D^{+}IS + \alpha 13c \cdot D^{+}C1 + \beta 22 \cdot D^{+}IF + kon \cdot O - (\alpha x1 + \beta 13c + \alpha 22 + koff) \cdot D^{+}O$$

$$\frac{dD^+C1}{dt} = \beta 13c \cdot D^+O + \alpha 33 \cdot D^+IF + \alpha 12 \cdot D^+C2 \ + kcon \cdot C1$$

$$- (\alpha 13c + \beta 33 + \beta 12 + kcoff) \cdot D^+C1$$

$$\frac{dD^+C2}{dt} = \beta 12 \cdot D^+C1 + \alpha 33 \cdot D^+IC2 + \alpha 11 \cdot D^+C3 \ + kcon \cdot C2$$

$$- (\alpha 12 + \beta 33 + \beta 11 + kcoff) \cdot D^+C2$$

$$\frac{dD^+C3}{dt} = \beta 11 \cdot D^+C2 + \alpha 33 \cdot D^+IC3 \ + kcon \cdot C3 - (\alpha 11 + \beta 33 + kcoff) \cdot D^+C3$$

$$\frac{dD^+IC3}{dt} = \beta 33 \cdot D^+C3 + \beta 11 \cdot D^+IC2 \ - (\alpha 33 + \alpha 11\ ) \cdot D^+IC3$$

$$\frac{dD^+IC2}{dt} = \alpha 11 \cdot D^+IC3 + \beta 33 \cdot D^+C2 + \beta 12 \cdot D^+IF \ - (\beta 11 + \alpha 33 + \alpha 12\ ) \cdot D^+IC2$$

$$\frac{dD^+IF}{dt} = \alpha 12 \cdot D^+IC2 + \beta 33 \cdot D^+C1 + \alpha 22 \cdot D^+O \ - (\beta 12 + \alpha 33 + \beta 22\ ) \cdot D^+IF$$

$$\frac{dD^+IS}{dt} = \alpha x1 \cdot D^+O - \beta x1 \cdot D^+IS$$

with rate constants

$$kon = kcon = [D^+]500\ M^{-1}$$

$$koff = kcoff = 500 * (318 \times 10^{-6}) e^{\frac{-0.7VF}{RT}}$$

$$\alpha x1 = (6.3992 \times 10^{-7})\alpha x$$

$$\beta x1 = 1.3511\beta x$$

$$\alpha 13c = (5.6974 \times 10^{-3})\alpha 13$$

$$\beta 13c = \frac{\beta 13 * kcon * koff * \alpha 13c}{kon * kcoff * \alpha 13}$$

$$\alpha 22 = (6.7067 \times 10^{-6})\alpha 2$$

$$\beta 22 = \frac{\alpha 13c * \alpha 22 * \alpha 33}{\beta 13c * \beta 33}$$

$$\beta 33 = (1.9698 \times 10^{-5})\beta 3$$

$$\alpha 33 = 3.2976\alpha 3$$

where $[D^+]$ is charged drug concentration in $M$, $R = 8314.472\ mJ/mol \cdot K$ is the gas constant, $F = 96485.3415\ C/mol$ is the Faraday constant, and $T$ is temperature in Kelvin.

Neutral drug-bound states:

$$\frac{dDO}{dt} = \beta x2 \cdot DIS + \alpha_13 \cdot DC1 + \beta_22 \cdot DIF + k_on \cdot O - (\alpha x2 + \beta_13 + \alpha_22 + k_off) \cdot DO$$

$$\frac{dDC1}{dt} = \beta_13 \cdot DO + \alpha_33 \cdot DIF + \alpha 12 \cdot DC2 + kc_on \cdot C1 - (\alpha_13 + \beta_33 + \beta 12 + k_off) \cdot DC1$$

$$\frac{dDC2}{dt} = \beta 12 \cdot DC1 + \alpha_33 \cdot DIC2 + \alpha 11 \cdot DC3 + kc_on \cdot C2 - (\alpha 12 + \beta_33 + \beta 11 + kc_off) \cdot DC2$$

$$\frac{dDC3}{dt} = \beta 11 \cdot DC2 + \alpha_33 \cdot DIC3 + kc_on \cdot C3 - (\alpha 11 + \beta_33 + kc_off) \cdot DC3$$

$$\frac{dDIC3}{dt} = \beta_33 \cdot DC3 + \beta 11 \cdot DIC2 + ki_on \cdot IC3 - (\alpha_33 + \alpha 11 + ki_off) \cdot DIC3$$

$$\frac{dDIC2}{dt} = \alpha 11 \cdot DIC3 + \beta_33 \cdot DC2 + \beta 12 \cdot DIF + ki_on \cdot IC2 - (\beta 11 + \alpha_33 + \alpha 12 + ki_off) \cdot DIC2$$

$$\frac{dDIF}{dt} = \alpha 12 \cdot DIC2 + \beta_33 \cdot DC1 + \alpha_22 \cdot DO + ki_on \cdot IF$$

$$- (\beta 12 + \alpha_33 + \beta_22 + ki_off) \cdot DIF$$

$$\frac{dDIS}{dt} = \alpha x2 \cdot DO + ki_on \cdot IS - (\beta x2 + ki_off) \cdot DIS$$

with rate constants

$$k_on = [D]500\ M^{-1}$$

$$k_off = 500(400 \times 10^{-6})$$

$$ki_on = k_on/2$$

$$ki_off = 500(3.4 \times 10^{-6})$$

$$kc_on = k_on/2$$

$$kc_off = 500(900 \times 10^{-6})$$

$$\alpha x2 = (1.3110 \times 10^{-1})\alpha x$$

$$\beta x2 = \frac{\beta x * k_on * \alpha x2 * ki_off}{\alpha x * ki_on * k_off}$$

$$\alpha_13 = (8.4559 \times 10)\alpha 13$$

$$\beta_13 = \frac{\beta 13 * kc_on * \alpha 13n * k_off}{kc_off * \alpha 13 * k_on}$$

$$\alpha_22 = (1.7084 \times 10^{-5})\alpha 2$$

$$\beta_22 = \frac{\alpha_33 * \alpha_13 * \alpha_22}{\beta_33\ * \beta_13}$$

$$\beta_33 = 4.8477\beta 3$$

$$\alpha_33 = \frac{ki_off * \alpha 3 * kc_on * \beta_{33}}{ki_on * kc_off * \beta 3}$$

where $[D]$ is neutral drug concentration in $M$.

## S7. Equations of the low-dimensional lidocaine-Na[+] channel interaction model

$$I_{Na} = \bar{g}_{Na} m^3 h (1-b)(V - E_{Na})$$

$$\frac{dm}{dt} = \alpha_m (1-m) - \beta_m m$$

$$\frac{dh}{dt} = \alpha_h (1-h) - \beta_h h$$

$$\frac{db}{dt} = [D] k_{on} (1-h)(1-b) - k_{off} b,$$

where maximal conductance per capacitance is $G_{Na} = \frac{\bar{g}_{Na}}{C_m} = 20\ nS/pF$ (units as in ten Tusscher et al. [1, 2] with $C_m$ being membrane capacitance), $k_{on} = 250$, $k_{off} = 1.7 \times 10^{-3}$, $[D]$ is the concentration of neutral lidocaine,

$$\alpha_m = 8.743 e^{\frac{V}{13.78}}, \beta_m = 0.1276 e^{\frac{V}{-23.25}}, \alpha_h = (1.187 \times 10^{-5}) e^{\frac{V}{-9.328}}, \beta_h = 2.723 e^{\frac{V}{14.91}}$$

at 22°C. The reversal potential $E_{Na} = \frac{RT}{F} \ln \frac{[Na^+]_{out}}{[Na^+]_{in}}$, where $R = 8314.472\ mJ/mol \cdot K$ is the gas constant, $F = 96485.3415\ C/mol$ is the Faraday constant, $T$ is temperature in Kelvin, and $[Na^+]_{out}$ and $[Na^+]_{in}$ are extracellular and intracellular $Na^+$ concentrations, respectively. Units of variables and parameters are $V$ in $mV$, $t$ in $ms$, $k_{on}$ in $M^{-1}ms^{-1}$, $k_{off}$ in $ms^{-1}$, $[D]$ in $M$ and transition rates ($\alpha$'s and $\beta$'s) in $ms^{-1}$. State transition rates are from fitting to drug-free voltage-clamp data in Results 3.1.1 of the main text, and drug binding rates $k_{on}$ and $k_{off}$ are taken from literature [3-5].

Using a $Q_{10}$ factor of 3, state transition rates were also adjusted for 37℃

$$\alpha_m = 45.43e^{\frac{V}{13.78}}, \beta_m = 0.6628e^{\frac{V}{-23.25}}, \alpha_h = (6.169 \times 10^{-5})e^{\frac{V}{-9.328}}, \beta_h = 14.15e^{\frac{V}{14.91}},$$

but lidocaine binding rates are unchanged (as is the case for neutral lidocaine binding rates in the Moreno et al. model [3]).